\documentclass[11pt]{article}
\usepackage{url}
\usepackage{amssymb,amscd,amstext,amsmath, amsthm, graphicx}
\usepackage{latexsym}
\usepackage{hyperref}
\usepackage[T2A]{fontenc}
\usepackage[cp1251]{inputenc}
\usepackage{floatflt}
\usepackage[mathscr]{eucal}
\usepackage{tikz}
\usetikzlibrary{arrows}
\usepackage{verbatim}
\usepackage{soul}
\usepackage{wrapfig}
\usepackage{makeidx} 
\usepackage{fancyhdr}
\usepackage{hyperref}
\usepackage{setspace}
 \usepackage{placeins}

\definecolor{linkcolor}{HTML}{000000} 
\definecolor{urlcolor}{HTML}{000000} 
\hypersetup{pdfstartview=Fit, linkcolor=linkcolor,urlcolor=urlcolor,
colorlinks=true, citecolor=linkcolor}

\newcommand*{\F}[1]{\ensuremath{\mathbb{F}_2^{#1}}}

\renewenvironment{abstract}{%
\begin{center}\begin{minipage}{0.9\textwidth}\begin{small}
\textbf{Abstract.}}
{\end{small}\par\noindent\end{minipage}\end{center}}

\title{\bf Mathematical methods in solutions of the problems from the Third International Students' Olympiad\\in Cryptography\footnote{
The work was supported by Russian Ministry of Science and Education under the 5-100 Excellence Programme, RMC NSU, and by the Russian Foundation for Basic Research (projects no. 15-07-01328, 17-41-543364).
}}
\author{N.~Tokareva$^{1,2}$,
    A.~Gorodilova$^{1,2}$,
    S.~Agievich$^{3}$,
    V.~Idrisova$^{1,2}$,
    N.~Kolomeec$^{1,2}$,\\
    A.~Kutsenko$^{1}$,
    A.~Oblaukhov$^{1}$,
    G.~Shushuev$^{1,2}$
	\\
\\
	{\small$^1$Novosibirsk State University, Novosibirsk, Russia}
\\
{\small$^2$Sobolev Institute of Mathematics, Novosibirsk, Russia}
\\
{\small$^3$Belarusian State University, Minsk, Belarus}
	\\
	{\small E-mail: {\tt nsucrypto@nsu.ru}}	
	}

\date{}

\begin{document}

\pagenumbering{arabic}

\maketitle

\begin{abstract}
The mathematical problems and their solutions of the Third International Students'
Olympiad in Cryptography NSUCRYPTO'2016 are presented. We consider
mathematical problems related to the construction of algebraic
immune vectorial Boolean functions and big Fermat numbers, problems
about secrete sharing schemes and pseudorandom binary sequences,
biometric cryptosystems and the blockchain technology, etc. Two
open problems in mathematical cryptography are also discussed and a solution for one of them proposed by a participant during the Olympiad is described. It was the first time in the Olympiad history.
\vspace{0.2cm}

\end{abstract}

\section{Introduction}

The Third International Students’ Olympiad in Cryptography --- NSUCRYPTO'2016 was held during November, 13~---~November, 21, 2016.
NSUCRYPTO is the unique cryptographic Olympiad containing
scientific mathematical problems for students and professionals from any country. Its
aim is to involve young researchers in solving curious and tough scientific problems of
modern cryptography. From the very beginning, the concept of the Olympiad was not
to focus on solving olympic tasks but on including unsolved research problems at the
intersection of mathematics and cryptography.

The Olympiad consisted of two independent Internet rounds. The First round (duration 4 hours 30 minutes) was individual and divided into two sections: A and B. Theoretical problems in mathematics of cryptography were offered to participants. The Second round (duration 1 week) was devoted to research and programming problems of cryptography solved in teams. Anyone who wanted to try his/her hand in solving cryptographic problems were able to become a participant. During the registration every participant had to choose corresponding category: ``School Student'' (for junior researchers: pupils and school students), ``University Student'' (for participants who were currently studying at universities) or ``Professional'' (for participants who had already completed education, are PhD students, or just wanted to be in the restriction-free category). The winners were awarded in each category separately. The language of the Olympiad is English. All information about organization and rules of the Olympiad can be found on the official website at \href{http://nsucrypto.nsu.ru}{\textcolor{blue}{\tt www.nsucrypto.nsu.ru}}.

In 2016 the geography of participants has expanded significantly. There were 420 participants from 24 countries: Russia (Novosibirsk, Moscow, Saint Petersburg, Yekaterinburg, Kazan, Saratov, Taganrog, Krasnoyarsk, Petrozavodsk, Perm, Chelyabinsk, Zelenograd, Tomsk, Korolev, Omsk, Ramenskoye, Yaroslavl, Novokuznetsk), Belarus (Minsk), Ukraine (Kiev, Kharkov, Zaporozhye), Kazakhstan (Astana, Almaty, Kaskelen), Kirghizia (Bishkek), Great Britain (Bristol), Bulgaria (Sofia), Germany (Berlin, Munich, Bochum, Witten), France (Paris), Luxembourg (Luxembourg), Hungary (Szeged), Sweden (Gothenburg), Switzerland (Zurich, Bern), Italy (Padova), Czech Republic (Prague), Estonia (Tartu), Spain (Barcelona) Canada (Edmonton), Iran (Tehran), South Africa (Cape Town), China (Beijing), Vietnam (Ho Chi Minh City, Saigon), Indonesia (Bandung), India (Kollam, Haydebarad).

Organizers of the Olympiad are: Novosibirsk State University, Sobolev Institute
of Mathematics (Novosibirsk), Tomsk State University, Belarusian State University
and University of Leuven (KU Leuven, Belgium).

In section \ref{problem-structure} we describe problem structure of the Olympiad according to sections and rounds. Section \ref{uns-problems} is devoted to unsolved problems formulated in NSUCRYPTO for all years since 2014, with attention to solutions proposed for two of them. Section \ref{problems} contains the conditions of all 16 mathematical problems of NSUCRYPTO'2016. Among them, there are both some amusing tasks based on historical ciphers as well as hard mathematical problems. We consider mathematical problems related to the construction of algebraically immune vectorial Boolean functions and big Fermat numbers, problems about secrete sharing schemes and pseudorandom binary sequences, biometric crypto\-systems and the blockchain technology, etc. Some unsolved problems are also discussed. In section \ref{solutions} we present solutions of all the problems with paying attention to solutions proposed by the participants.
The lists of the winners are given in section \ref{winners}.

Mathematical problems of the previous International Olympiads
NSUCRYPTO'2015 and NSUCRYPTO'2016 can be found in \cite{paper-pdm} and \cite{paper-cryptologia} respectively.

\section{Problem structure of the Olympiad}
\label{problem-structure}

There were 16 problems on the Olympiad. Some of them were included
in both rounds (Tables\;\ref{Probl-A},\,\ref{Probl-B},\,\ref{Probl-Second}).
Thus, section A (school section) of the first round consisted of 6 problems, whereas section B
(student section) contained 7 problems. Three problems were common for both sections.
In the table you can see the highest scores one could get in case of solving the problem.

\begin{table}[h]
\centering\footnotesize
\caption{{\bf Problems of the first round (A --- school section)}}
\label{Probl-A}
\medskip
\begin{tabular}{|c|l|c|}
  \hline
  N & Problem title & Maximum scores \\
  \hline
  1 & Cipher from the pieces & 4 \\
  2 & Get an access & 4 \\
  3 & Find the key & 4 \\
  4 & Labyrinth & 4 \\
  5 & System of equations & 4 \\
  6 & Biometric pin-code & 4 \\
  \hline
\end{tabular}
\end{table}

\begin{table}[h]
\centering\footnotesize
\caption{{\bf Problems of the first round (B --- student section)}}
\label{Probl-B}
\medskip
\begin{tabular}{|c|l|c|}
  \hline
  N & Problem title & Maximum scores \\
  \hline
  1 & Cipher from the pieces & 4 \\
  2 & Labyrinth & 4 \\
  3 & Quadratic functions & 8 \\
  4 & System of equations & 4 \\
  5 & Biometric key & 6 \\
  6 & Secret sharing & 6 \\
  7 & Protocol & 6 \\
  \hline
\end{tabular}
\end{table}

The second round was composed of 12 problems; they were common for all the
participants. Two of the problems presented on the second round were marked as
unsolved (awarded special prizes from the Program Committee).

\begin{table}[h]
\centering\footnotesize
\caption{{\bf Problems of the second round}}
\label{Probl-Second}
\medskip
\begin{tabular}{|c|l|c|}
  \hline
  N & Problem title & Maximum scores \\
  \hline
  1 & Algebraic immunity & Unsolved \\
  2 & Zerosum at AES & 8 \\
  3 & Latin square & 6 \\
  4 & Nsucoin & 10 \\
  5 & Metrical cryptosystem & 6 \\
  6 & Quadratic functions & 8 \\
  7 & Secret sharing & 6 \\
  8 & Biometric key & 6 \\
  9 & Protocol & 6 \\
  10 & Find the key & 4 \\
  11 & Labyrinth & 4 \\
  12 & Big Fermat numbers & Unsolved \\
  \hline
\end{tabular}
\end{table}

\section{Unsolved problems of NSUCRYPTO since the origin}
\label{uns-problems}

In this section we shortly present all unsolved problems stated in the history of NSUCRYPTO (Table \ref{Tab-unsolved}), where we mention also the current status of all the problems. The formulations of the problems can be found in \cite{paper-pdm} (2014), \cite{paper-cryptologia} (2015) as well as the current paper (2016).

\begin{table}[h]
\centering\footnotesize
\caption{{\bf Unsolved Problems of NSUCRYPTO}}
\label{Tab-unsolved}
\medskip
\begin{tabular}{|c|c|l|l|}
  \hline
  N & Year & Problem title & Status \\
  \hline
  1 & 2014 & Watermarking cipher & Unsolved \\
  2 & 2014 & APN permutation     & Unsolved \\
  3 & 2014 & Super S-box         & Unsolved \\

  4 & 2015 & A secret sharing    & Partially SOLVED in \cite{17-Titov} \\
  5 & 2015 & Hypothesis          & Unsolved \\

  6 & 2016 & Algebraic immunity  & SOLVED during the Olympiad \\
  7 & 2016 & Big Fermat numbers  & Unsolved \\
  \hline
\end{tabular}
\end{table}

The Olympiad NSUCRYPTO-2016 became particular since there was the first time when an unsolved problem stated was successfully solved by a participant during the Olympiad. Alexey Udovenko (University of Luxembourg) was able to find a solution to the problem ``Algebraic immunity'' (see section \ref{AI}).

Moreover, we are also very pleased to say that the unsolved problem ``A secret sharing'' of NSUCRYPTO-2015 was also partially solved! In \cite{17-Titov} Kristina Geut, Konstantin Kirienko, Prokhor Kirienko, Roman Taskin, Sergey Titov (Ural State University of Railway Transport, Yekaterinburg) found a solution for the problem in the case of even dimension (the problem remains open for odd dimension).

The current status of unsolved problems from NSUCRYPTO of all years can be found at \href{http://nsucrypto.nsu.ru/unsolved-problems/}{\textcolor{blue}{\tt http://nsucrypto.nsu.ru/unsolved-problems/}}.
Everyone is welcome to propose a solution to any problem stated. Please, send you ideas to {\textcolor{blue}{\tt nsucrypto@nsu.ru}}.

\section{Problems}
\label{problems}
In this section we formulate all the problems of the Olympiad.

\subsection{Problem ``Cipher from the pieces''}

Recover the original message, splitting the figure into equal pieces such that each color occurs once in every piece.

\begin{center}
\includegraphics[scale=0.3]{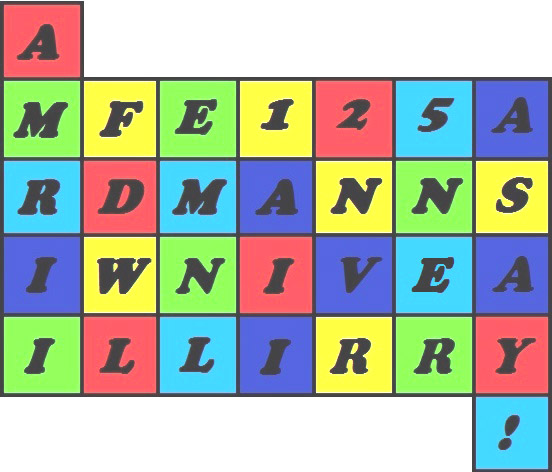}
\end{center}

\subsection{Problem ``Get an access''}

To get an access to the safe one should put 20 non-negative integers in the following cells (Fig.\;\ref{probl2}). The safe will be opened if and only if the sum of any two numbers is even number $k$, such that $4 \leqslant k \leqslant 8$, and each possible sum occurs at least once. Find the sum of all these numbers.

\begin{center}
\includegraphics[scale=0.3]{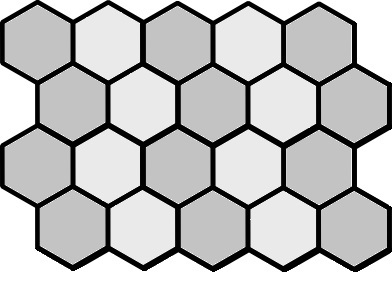}
\end{center}

\subsection{Problem ``Find the key''}

The key of a cipher is the set of positive integers $a$, $b$, $c$, $d$, $e$, $f$, $g$, such that the following relation holds:
\begin{center}
$a^3+b^3+c^3+d^3+e^3+f^3+g^3=2016^{2017}.$
\end{center}

Find the key!

\subsection{Problem ``Labyrinth''}

Read the message hidden in the labyrinth!

\begin{center}
\includegraphics[scale=0.3]{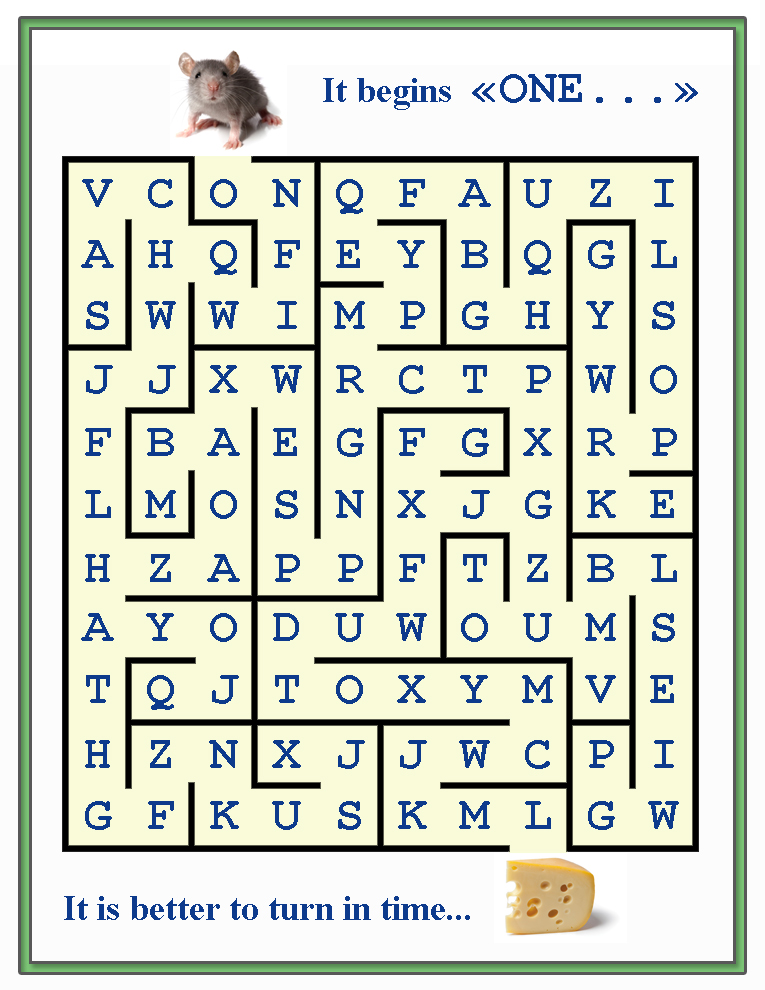}
\end{center}

\subsection{Problem ``System of equations''}

Analyzing a cipher Caroline gets the following system of equations in binary variables $x_1,x_2,\ldots,x_{16}\in\{0,1\}$ that represent the unknown bits of the secrete key:

\begin{equation*}
 \begin{cases}
   x_1x_3\oplus x_2x_4=x_5-x_6, \\
   x_{14}\oplus x_{11}=x_{12}\oplus x_{13}\oplus x_{14}\oplus x_{15}\oplus x_{16}, \\
   (x_8+x_9+x_7)^2=2(x_6+x_{11}+x_{10}), \\
   x_{13}x_{11}\oplus x_{12}x_{14}=-(x_{16}-x_{15}), \\
   x_5x_1x_6=x_4x_2x_3, \\
   x_{11}\oplus x_8\oplus x_7=x_{10}\oplus x_{6}, \\
   x_6x_{11}x_{10}\oplus x_7x_9x_8=0, \\
   \left(\frac{x_{12}+x_{14}+x_{13}}{\sqrt{2}}\right)^2-x_{15}=x_{16}+x_{11}, \\
   x_1\oplus x_6=x_5\oplus x_3\oplus x_2, \\
   x_6x_8\oplus x_9x_7=x_{10}-x_{11}, \\
   2(x_5+x_1+x_6)=(x_4+x_3+x_2)^2, \\
   x_{11}x_{13}x_{12}=x_{15}x_{14}x_{16}. \\
 \end{cases}
\end{equation*}
\medskip

Help Caroline to find the all possible keys!

\medskip
{\bf Remark.}
If you do it in analytic way (without computer calculations) you get twice more scores.

\subsection{Problem ``Biometric pin-code''}

{\small Iris is one of the most reliable biometric characteristics
of a human. While measuring let us take 16-bit vector from the
biometric image of an iris. As in reality, we suppose that two
16-bit biometric images of {\it the same human} can differ not more
than by $10$--$20\%$, while biometric images of {\it different
people} have differences at least $40$--$60\%$.}

\begin{center}
\includegraphics[scale=0.4]{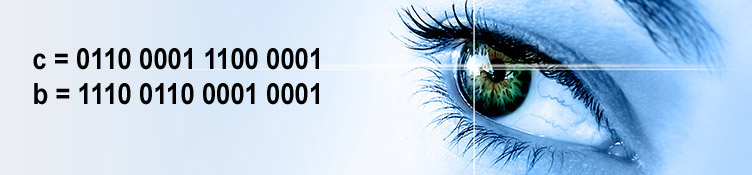}
\end{center}

{\small Let a key $k$ be an arbitrary 5-bit vector. We suppose that
the key is a pin-code that should be used in order to get an access to
the bank account of a client.

To avoid situation when malefactor can steal the key of a some
client and then be able to get an access to his account, the bank
decided to combine usage of the key with biometric authentication of
a client by iris-code. The following scheme of covering the key with
biometric data was proposed:}

1) on registration of a client take 16-bit biometric image
$b_{template}$ of his iris;

2) extend 5-bit key $k$ to 16-bit string $s$ using Hadamard
encoding, i.\,e. if $k=(k_1,\ldots ,k_5)$, where $k_i\in \{0, 1\}$,
then $s$ is the vector of values of the Boolean function
$f(x_1,\ldots, x_4) = k_1x_1 \oplus \ldots \oplus k_4x_4 \oplus
k_5$, where $\oplus$ is summing modulo $2$;

3) save the vector $c = b_{template} \oplus s$ on the smart-card and
give it to the client. A vector $c$ is called {\it biometrically
encrypted key}.

To get an access to his account a client should

1) take a new 16-bit biometric image $b$ of his iris;

2) using information from the smart-card count 16-bit vector $s'$ as
$s' = b\oplus c$;

3) decode $s'$ to the 5-bit vector $k'$ using Hadamard decoding
procedure.

Then the bank system checks: if $k'=k$ then the client is
authenticated and the key is correct; hence bank provides an access
to the account of this client. Otherwise, if $k'\neq k$ then bank
signals about an attempt to get illegal access to the bank account.

\medskip
{\bf The problem.} Find the 5-bit $k$ of Alice if you know her
smart-card data $c$ and a new biometric image $b$ (both are given on
the picture).

\medskip
{\bf Remark.}  Vector of values of a Boolean function $f$ in
$4$ variables is a binary vector
$(f(x^0),f(x^1)$, $\ldots, f(x^{15}))$ of length $16$, where $x^0 = (0,
0, 0, 0)$, $x^1 = (0, 0, 0, 1)$, $\ldots$, $x^{15} = (1, 1 , 1, 1)$,
ordered by lexicographical order; for, example, vector of values of
the function $f(x_1,x_2,x_3,x_4) = x_3\oplus x_4 \oplus 1$ is equal
to $(1010101010101010)$.

\subsection{Problem ``Quadratic functions''}

Alice and Bob are going to use the following pseudo\-random binary sequence
$u = \{u_i\}$, $u_i \in \F{}$:

\begin{itemize}
\item $u_1, \ldots, u_n$ are initial values;
	\item $u_{i + n} = f(u_{i}, u_{i + 1}, \ldots, u_{i + n - 1})$, where
$$
	f \in Q_{n} = \{ a_0 \oplus \bigoplus_{i = 1}^{n}a_ix_i \oplus \bigoplus_{1 \leq i < j \leq n}{a_{ij}x_i x_j  \ | \ a_0, a_i, a_{ij} \in \mathbb{F}_2}\}.
$$
\end{itemize}

Suppose that you have intercepted the elements  $u_{t}, u_{t + 1}, \ldots, u_{t + k - 1}$ of a sequence for some $t$. Is it possible to uniquely reconstruct the elements $$u_{t + k}, u_{t + k + 1},  u_{t + k + 2}, \ldots$$  provided $k \leqslant c n$, where $c$ is a constant independent on $n$?

\subsection{Problem ``Biometric key''}

Iris is one of the most reliable biometric characteristics of a
human. While measuring let us take 128-bit biometric image of an
iris. As in reality, we suppose that two 128-bit biometric images of
{\it the same human} can differ not more than by $10$--$20\%$, while
biometric images of {\it different people} have differences at least
$40$--$60\%$.
\begin{center}
\includegraphics[scale=0.4]{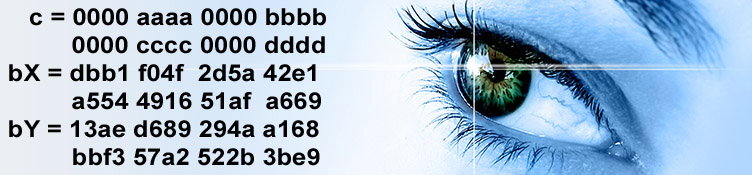}
\end{center}

Let a key $k$ be an arbitrary 8-bit vector. It can be represented in
hexadecimal notation. For example, ${\tt e2} = 1110 0010$. We
suppose that the key is a pin-code that should be used in order to
get access to the bank account of a client.

To avoid situation when malefactor can steal the key of a some
client and then be able to get an access to his account, the bank
decided to combine usage of the key with biometric authentication of
a client by iris-code. The following scheme of covering the key with
biometric data was proposed:

1) on registration of a client take 128-bit biometric image
$b_{template}$ of his iris;

2) extend 8-bit key $k$ to 128-bit string $s$ using Hadamard
encoding, i.\,e. if $k=(k_1,\ldots ,k_8)$, where $k_i\in
\mathbb{F}_2$, then $s$ is the vector of values of the Boolean
function $f(x_1,\ldots, x_7) = k_1x_1 \oplus \ldots \oplus k_7x_7
\oplus k_8$, where $\oplus$ is summing modulo $2$;

3) save the vector $c = b_{template} \oplus s$ on the smart-card and
give it to the client. A vector $c$ is called {\it biometrically
encrypted key}.

To get an access to his account a client should

1) take a new 128-bit biometric image $b$ of his iris;

2) using information from the smart-card count 128-bit vector $s'$
as $s' = b\oplus c$;

3) decode $s'$ to 8-bit vector $k'$ using Hadamard decoding
procedure.

Then the bank system checks: if $k'=k$ then the client is
authenticated and the key is correct; hence bank provides an access
to the account of this client. Otherwise, if $k'\neq k$ then bank
signals about an attempt to get illegal access to the bank account.

\medskip
{\bf The problem.} One day a person, say X, came to the bank and
tried to get an access to the bank account of Alice using the
smart-card. This may be noticed that person X was in hurry and may
be a little bit nervous. Suddenly, another person, say Y, appeared
in the bank and declared loudly: ``Please stop any operation! I am
Alice! My smart-card was stolen.''

Bank clerk, say Claude, stopped all operations. In order to solve
the situation he took new biometric images $b^X$ and $b^Y$ of
persons X and Y respectively, and with smart-card containing vector
$c$ leaved his post for consultations with bank specialists.

When Claude came back, he already knew who was Alice. He wanted to
stop the other person and call to police but that person has already
disappeared. So, can you solve this problem too? Who was real Alice?
Determine her 8-bit key $k$. You can use the data $b^X$, $b^Y$ and
$c$ presented on the picture. It is known also that the key of Alice
contains odd number of ones.

\medskip
{\bf Remark.}  The vector of values of a Boolean function $f$
in $n$ variables is a binary vector $(f(x^0),f(x^1),\ldots,
f(x^{2^n-1}))$ of length $2^n$, where $x^0 = (0, \ldots, 0, 0)$,
$x^1 = (0, \ldots , 0, 1)$, $\ldots$, $x^{2^n-1} = (1, \ldots ,
1,1)$, ordered by lexicographical order; for example, the vector of
values of the function $f(x_1,x_2) = x_1\oplus x_2 \oplus 1$ is
equal to $(1001) = {\tt 9}$. The vector of values of the function
$f(x_1,\ldots, x_8) = x_1\oplus x_2 \oplus 1$ is {\tt ffff ffff 0000
0000 0000 0000 ffff ffff}.

\subsection{Problem ``Secret sharing''}

Alena, Boris and Sergey developed the following secret sharing scheme to share a password $P \in \F{32}$ into three parts to collectively manage money through online banking.

\begin{itemize}
	\item Vectors $v_i^a, v_i^b, v_i^s \in \F{32}$ and values $c_i^a, c_i^b, c_i^s \in \F{}$ are randomly generated for all $i = 1, \ldots, 32$.
	\item Vectors $v_i^a, v_i^b, v_i^s$ are known to all participants of the scheme.
	\item Values $c_i^a, c_i^b, c_i^s \in \F{}$ are known only to Alena, Boris and Sergey respectively.
\item Then the secret password $P$ is calculated by the rule
$$
	P = \bigoplus_{i = 1}^{32}{c_i^a v_i^a} \oplus \bigoplus_{i = 1}^{32}{c_i^b v_i^b} \oplus \bigoplus_{i = 1}^{32}{c_i^s v_i^s}.
$$
\end{itemize}

What is the probability that Alena and Boris together can not get any information about the password $P$? What is the probability that they are able without Sergey to get a guaranteed access to online banking using not more than 23 attempts?

\subsection{Problem ``Protocol''}

Alena and Boris developed a new protocol for establishing session keys. It consists of the following three steps:

\medskip

1. The system has a common prime modulus $p$ and a generator $g$. Alena and Boris have their own private keys $\alpha_a \in \mathbb{Z}_{p-1}$, $\alpha_b \in \mathbb{Z}_{p-1}$ and corresponding public keys $P_a=g^{\alpha_a}\!\mod{p}$, $P_b=g^{\alpha_b}\!\mod{p}$.

\medskip

2. To establish a session key Alena generates a random number $R_a\in \mathbb{Z}_{p-1}$, computes $X_a=(\alpha_a + R_a)\!\mod{(p-1)}$ and sends it to Boris. Then Boris generates his random number $R_b$, computes $X_b$ in the same way as Alena and sends it back to~her.

\medskip

3. Alena computes the session key in the following way:

$$K_{a,b}= (g^{X_b}P_b^{-1})^{R_a}\!\mod{p}.$$

Bob computes the session key in the following way:

$$K_{b,a}= (g^{X_a}P_a^{-1})^{R_b}\!\mod{p}.$$

\medskip
How can an attacker Evgeniy compute any future session key between Alena and Boris, if he steals the only one session key $K_{a,b}$?

\subsection{Problem ``Zerosum at AES''}

Let $\texttt{AES}_0$ be a mapping that represents the algorithm AES-256 with the all-zero key. Let $X_1,\ldots,X_{128}\in\F{128}$ be pairwise different vectors such that
$$
\bigoplus_{i=1}^{128} X_i=\bigoplus_{i=1}^{128}\texttt{AES}_0(X_i).
$$
\begin{enumerate}
	\item Propose an effective algorithm to find an example of such vectors $X_1,\ldots,X_{128}$.
	\item Provide an example of $X_1,\ldots,X_{128}$.
\end{enumerate}

\subsection{Problem ``Latin square''}

Alice has registered on Bob's server. During the registration Alice got the secret key that is represented as a latin square of order $10$. A latin square is a $10\times10$ matrix filled with integers $0, 1, \ldots, 9$, each occurring exactly once in each row and exactly once in each column.

To get an access to Bob's resources Alice authenticates by the following algorithm:

\begin{enumerate}
	\item Bob sends to Alice a decimal number $abcd$, where $a, b, c, d \in \{0, 1, \ldots, 9\}$ and $a\neq b$, $b\neq c$, $c\neq d$.
	\item Alice performs three actions.
	\begin{itemize}	
	
		\item At first she finds the integer~$t_1$ standing at the intersection of the row~$(a + 1)$ and the column~$(b + 1)$.
		\item Then she finds~$t_2$ standing at the intersection of the row~$(t_1 + 1)$ and the column~$(c + 1)$.
		
		\item Finally, Alice finds~$t_3$ standing at the intersection of the row~$(t_2 + 1)$ and the column~$(d + 1)$.
	\end{itemize}
	\item Alice sends to Bob the integer~$t_3$.
	\item Bob performs the same actions and verifies Alice's answer.

	\item Steps 1-4 are repeated several times. In case of success Bob recognizes that Alice knows the secret latin square.
\end{enumerate}

\noindent Find Alice's secret key if you can get the answer $t_3$ for
any your correct input request $abcd$~\href{http://nsucrypto.nsu.ru/archive/2016r2/task3}{\textcolor{blue}{here}}.

\subsection{Problem ``\textsl{nsucoin}''}

Alice, Bob, Caroline and Daniel are using a digital payment system {\bf \textsl{nsucoin}} to buy from each other different sorts of flowers. Alice sells only chamomiles, Bob --- only tulips, Caroline --- only gerberas and Daniel --- only roses.
At the beginning each person has 5 flowers. The cost of each flower is 2 coins.

{\bf Transactions} are used to make purchases by transferring coins in the system \textsl{nsucoin}.
Each transaction involves two different users (the seller $A$ and the buyer $B$) and distributes a certain amount of coins $S$ between $A$ and $B$, say $S = S_A + S_B$. The value $S$ is equal to the sum of all the coins received by the buyer in the indicated $k$ transactions, $1\leqslant k \leqslant 2$. We will say that the current transaction is {\it based} on these~$k$ transactions. The value $S_A$ is the amount of coins that the buyer pays the seller for his product, $S_A>0$; the value $S_B$ is the rest of available amount of coins $S$ that returns to buyer (in further transactions $B$ can spend these coins).
At the same time, coins received by users in each transaction can not be distributed more than once in other transactions.

\begin{center}
\includegraphics[scale=0.3]{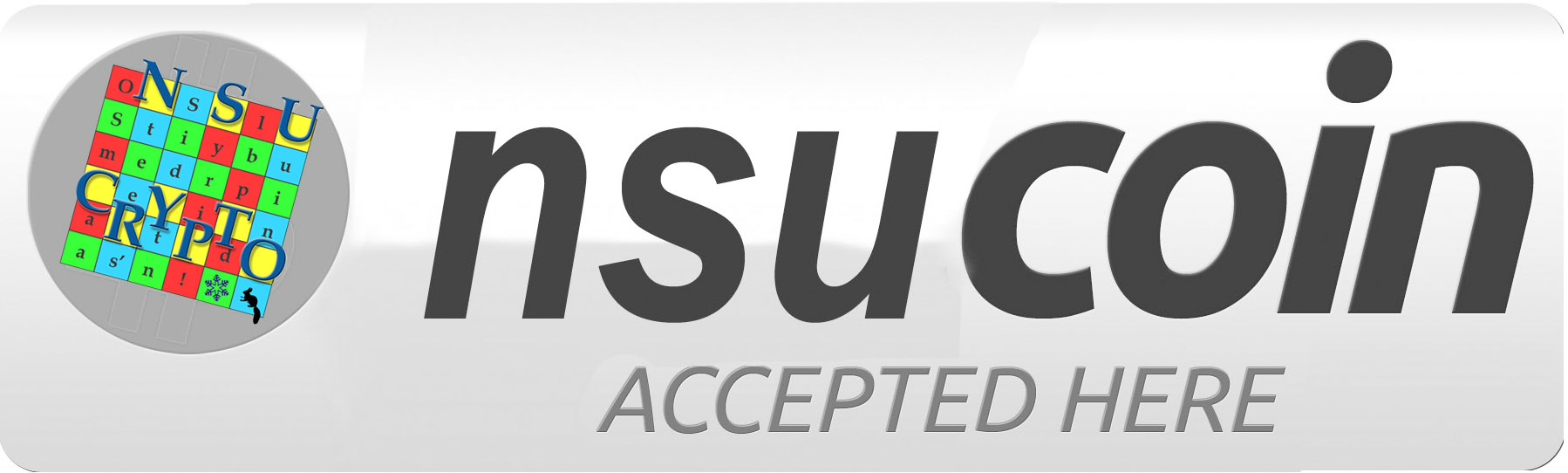}
\end{center}

In order for transactions to be valid they must be verified. To do this {\bf block chain} is used. Each block verifies from 1 to 4 transactions. Each transaction to be verified can be based on already verified transactions and transactions based on verified transactions.

There are 4 {\it special} transactions. Each of them brings 10 coins to one user. These transactions do not based on other transactions. The first block verifies all special transactions.

Define what bouquet Alice can make from the flowers she has
if the last block in chain is the following string (hash of this block in {\tt 00004558}):
\begin{center}
{\tt height:2;prevHash:0000593b;ctxHash:8fef76cb;nonce:17052}
\end{center}

\noindent{\bf Technical description of \textsl{nsucoin}.}
\medskip

{\bf $\bullet$ Transactions.}
Transaction is given by the string {\tt transaction} of the following format:

\medskip
{\tt transaction = ``txHash:\{hashValue\};\{transactionInfo\}''}

{\tt hashValue = Hash(\{transactionInfo\})}

{\tt transactionInfo = ``inputTx:\{Tx\};\{sellerInfo\};\{buyerInfo\}''}

{\tt Tx = ``\{Tx1\}'' or ``\{Tx1,Tx2\}''}

{\tt sellerInfo = ``value1:\{V1\};pubKey1:\{PK1\};sign1:\{S1\}''}

{\tt buyerInfo = ``value2:\{V2\};pubKey2:\{PK2\};sign2:\{S2\}''}

\medskip

Here {\tt Tx1, Tx2} are values of the field {\tt txHash} of transactions which the current transaction based on. {\tt V{\it i}} is a non-negative integer that is equal to the amount of coins received by the user with public key {\tt PK{\it i}}, $0\leqslant{\text{\tt V{\it i}}}\leqslant10$, {\tt V1}$\neq0$. Digital signature
\begin{center}{\tt S{\it i}} = {\tt DecToHexStr}({\tt Signature}({\tt Key2},{\tt StrToByteDec}({\tt Hash}({\tt Tx1+Tx2+PK{\it i}})))),\end{center} where {\tt +} is concatenation operation of strings. {\tt Key2} is private key of buyer.

In the special transactions fields {\tt inputTx}, {\tt sign1} are empty and there is no {\tt buyerInfo}. For example, one of the special transactions is the following:
\begin{center}
{\tt txHash:1a497b59;inputTx:;value1:10;pubKey1:11;sign1:}
\end{center}

\medskip
{\bf $\bullet$ Block chain.} Each block is given by the string {\tt block} of the following format:

\medskip

{\tt block = ``height:\{Height\};prevHash:\{PrHash\};ctxHash:\{CTxHash\};nonce:\{Nonce\}''}

\medskip

Here {\tt Height} is the block number in a chain, the first block has number 0. {\tt PrHash} is hash of block with number $\text{\tt Height}-1$. {\tt CTxHash} is hash of concatenation of all the {\tt TxHash} of transactions verified by this block. {\tt Nonce} is the minimal number from 0 to 40000 such that block has hash of the form {\tt 0000\#\#\#\#}.

Let {\tt PrHash} = {\tt 00000000} for the first block.

\medskip
{\bf $\bullet$ Hash function.}
 {\tt Hash} is calculated as reduced MD5: the result of hashing is the first 4 bytes of standard MD5 represented as a string. For example, {\tt Hash}({\tt ``teststring''}) = {\tt ``d67c5cbf''}, {\tt Hash}({\tt ``1a497b5917''}) = {\tt ``e0b9e4a8''}.

\medskip
{\bf $\bullet$ Digital signature.}
{\tt Signature}$(key,message)$ is RSA digital signature with $n$ of order 64 bits, $n= 9101050456842973679$. Public exponents {\tt PK} of users are the following~(Table\;\ref{Tab-Pubkeys}).

\begin{table}[ht]
\centering\footnotesize
\caption{{\bf Public keys}}
\label{Tab-Pubkeys}
\medskip
\begin{tabular}{|l||l|l|l|l|}
\hline
User & Alice & Bob & Caroline & Daniel  \\
\hline
{\tt PK} & 11 & 17 & 199 & 5  \\
\hline\end{tabular}
\end{table}
For example,

{\tt Signature}(2482104668331363539, 7291435795363422520) = 7538508415239841520.

\medskip
{\bf $\bullet$ Additional functions.}
{\tt StrToByteDec} decodes a string to bytes that are considered as a number.
Given a number {\tt DecToHexStr} returns a string that is equal to the hexadecimal representation of this number.
For example, {\tt StrToByteDec}({\tt ``e0b9e4a8''}) = 7291435795363422520 and {\tt DecToHexStr}(7538508415239841520) = {\tt ``689e297682a9e6f0''}.

Strings are given in UTF-8.
\medskip

\noindent{\bf Examples of a transaction and a block.}
\medskip

 $\bullet$ Suppose that Alice are buying from Bob 2 tulips. So, she must pay him 4 coins. The transaction of this operation, provided that Alice gets 10 coin in the transaction with hash {\tt 1a497b59}, is
\begin{center}
{\tt txHash:98e93fd5;inputTx:1a497b59;value1:4;pubKey1:17;sign1:689e297682a9e6f0;\\value2:6;pubKey2:11;sign2:fec9245898b829c
}
\end{center}

$\bullet$ The block on height 2 verifies transactions with hash values (values of {\tt txHash})  {\tt 98e93fd5}, {\tt c16d8b22}, {\tt b782c145} and {\tt e1e2c554}, provided that hash of the block on height 1 is {\tt 00003cc3}, is the following:
\begin{center}
{\tt
height:2;prevHash:00003cc3;ctxHash:9f8333d4;nonce:25181
}
\end{center}
Hash of this block is {\tt 0000642a}.

\subsection{Problem ``Metrical cryptosystem''}


Alice and Bob exchange messages using the following cryptosystem.
Let $\mathbb{F}_2^n$ be an $n$-dimensional vector space over the
field $\mathbb{F}_2=\{0,1\}$. Alice has a set
$A\subseteq\mathbb{F}_2^n$ and Bob has a set
$B\subseteq\mathbb{F}_2^n$ such that both $A$ and $B$ are metrically
regular sets and they are metrical complements of each other. Let
$d$ be the Hamming distance between $A$ and $B$. To send some number
$a$ ($0 \leqslant a \leqslant d$) Alice chooses some vector $x\in
\mathbb{F}_2^n$ at distance $a$ from the set $A$ and sends this
vector to Bob. To obtain the number that Alice has sent Bob
calculates the distance $b$ from $x$ to the set $B$ and concludes
that the initial number $a$ is equal to $d-b$.

Is this cryptosystem correct? In other words, does Bob correctly decrypt all sent
messages, regardless of initial sets $A$, $B$ satisfying given
conditions and of the choice of vector $x$?

\medskip
\textbf{Remark I.} Recall several definitions and notions. The {\it
Hamming distance} $d(x,y)$ between vectors $x$ and $y$ is the number
of coordinates in which these vectors differ. Distance from vector
$y\in\mathbb{F}_2^n$ to the set $X\subseteq\mathbb{F}_2^n$ is
defined as $d(y,X)=\min_{x\in X} d(y,x)$. The {\it metrical
complement} of a set $X\subseteq\mathbb{F}_2^n$ (denoted by
$\widehat{X}$) is the set of all vectors $y\in\mathbb{F}_2^n$ at
maximum possible distance from $X$ (this maximum distance is also
known as {\it covering radius} of a set). A set
$X\subseteq\mathbb{F}_2^n$ is called {\it metrically regular}, if
its second metrical complement $\widehat{\widehat{X}}$ coincides
with $X$.

\medskip
\textbf{Remark II.} Let us consider several examples: {\small
\begin{itemize}
  \item {Let $X$ consist of a single vector $x\in \mathbb{F}_2^n$. It is easy to see that $\widehat{X}=\{x\oplus \textbf{1}\}$,
where $\textbf{1}$ is the all-ones vector, and therefore $\widehat{\widehat{X}}=\{x\oplus \textbf{1}\oplus \textbf{1}\}=\{x\}=X$, so $X$ is a metrically regular set; it is also easy to see that cryptosystem based on $A=\{x\}$, $B=\{x\oplus\textbf{1}\}$ is correct;}
  \item {Let $Y$ be a ball of radius $r>0$ centered at $x$: $Y=B(r,x)=\{y\in \mathbb{F}_2^n:d(x,y)\leqslant r\}$. You can verify that $\widehat{Y}=\{x\oplus \textbf{1}\}$, but $\widehat{\widehat{Y}}=\{x\}\neq Y$, and $Y$ is not metrically regular;}
  \item {Let $X$ be an arbitrary subset of $\mathbb{F}_2^n$. Then, if we denote $X_0:=X$, $X_{k+1}=\widehat{X_k}$ for $k\geqslant 0$, there exists a number $M$ such that $X_m$ is a metrically regular set for all $m > M$. You can prove this fact as a small exercise, or simply use it in your solution.}
\end{itemize}}

\subsection{Problem ``Algebraic immunity'' (Unsolved)}

A mapping $F$ from $\F{n}$ to $\F{m}$ is called a {\it vectorial Boolean function} (recall
that $\F{n}$ is the vector space of all binary vectors of length $n$). If $m=1$ then $F$ is a {\it Boolean function} in $n$ variables. A {\it component function} $F_v$ of $F$ is a Boolean function defined by a vector $v\in\mathbb{F}_2^m$ as follows $F_v = \langle v, F\rangle = v_1 f_1 \oplus \ldots \oplus v_m f_m$, where $f_1,\ldots,f_m$ are coordinate functions of $F$.
A function $F$ has its unique {\it algebraic normal form} (ANF)
$$F(x) = \bigoplus_{I\in \mathcal{P}(N)} a_{I} \big(\prod_{i\in I} x_i \big),$$
where $\mathcal{P}(N)$ is the power set of $N=\{1,\ldots,n\}$ and $a_I$ belongs to $\mathbb{F}_2^m$. Here $\oplus$ denotes the coordinate-wise sum of vectors modulo 2. The {\it algebraic degree} of $F$ is the degree of its ANF: ${\deg}(F) = \max\{|I|:\ a_I\neq(0,\ldots,0),\ I\in\mathcal{P}(N)\}$.

\medskip
{\bf Algebraic immunity} $AI(f)$ of a Boolean function $f$ is the minimal algebraic degree of a Boolean function $g$, $g\not\equiv 0$, such that $fg\equiv 0$ or $(f\oplus 1)g \equiv 0$. The notion was introduced by W.\,Meier, E.\,Pasalic, C.\,Carlet in 2004.

\medskip
{\bf The tight upper bound of $AI(f)$}.
It is wellknown that $AI(f)\leqslant \lceil\frac{n}{2}\rceil$, where $\lceil x\rceil$ is the ceiling function of number $x$. There  exist functions with $AI(f)=\lceil\frac{n}{2}\rceil$ for any $n$.

\medskip
{\bf Component algebraic immunity} $AI_{comp}(F)$ of a function from $\F{n}$ to $\F{m}$ is defined as the minimal algebraic immunity of its component functions $F_v$, $v\neq (0,\ldots,0)$.
Component algebraic immunity was considered by C.\,Carlet in 2009.
It is easy to see that $AI_{comp}(F)$ is also upper bounded by $\lceil\frac{n}{2}\rceil$.

\medskip
{\bf The problem.} What is the tight upper bound of component algebraic immunity? For all possible combination of $n$ and $m$, $m\leqslant n\leqslant 4$, vectorial Boolean functions with $AI_{comp}(F)=\lceil\frac{n}{2}\rceil$ exist.

\medskip
Construct $F: \mathbb{F}_2^5\to \mathbb{F}_2^5$ with maximum
possible algebraic component immunity $3$ or prove that it does not exist.

\subsection{Problem ``Big Fermat numbers'' (Unsolved)}

It is known that constructing big prime numbers is very actual and
complicated problem interesting for cryptographic applications. One
of the popular way to find them is... to guess! For example to guess
them between numbers of some special form. For checking there are
Mersenne numbers $2^k - 1$, Fermat numbers $F_k = 2^{2^k}+1$ for
nonnegative integer $k$, etc.

Let us concentrate our attention on Fermat's numbers.

It is known that Fermat numbers $F_0 = 3$, $F_1 = 5$, $F_2 = 17$,
$F_3 = 257$, $F_4 = 65 537$ are prime. But the number $F_5 = 4\,
284\, 967\, 297 = 641\cdot 6\, 700\, 417$ is already composite as
was proven by L.\,Euler in XVIII.

For now it is known that all Fermat numbers, where $k=5,\ldots,
32$, are composite and there is the hypothesis that every Fermat
number $F_k$, where $k\geqslant 5$ is composite.

Could you prove that for any big number $N$ there exists a composite
Fermat number $F_k$ such that $F_K >N$?

\section{Solutions of the problems}
\label{solutions}

In this section we present solutions of the problems with paying attention to solutions proposed by the participants (right/wrong and beautiful).

\subsection{Problem ``Cipher from the pieces''}

{\bf Solution.} The only way to split this figure is the following.
\begin{center}
\includegraphics[scale=0.3]{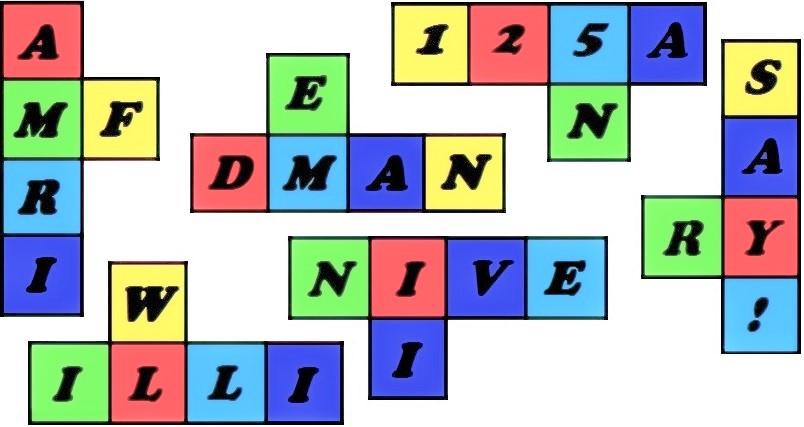}
\end{center}

Then we need to arrange these pieces and read letters horizontally. In this way we can obtain the text  <<{\tt WILLI AMFRI EDMAN 125AN NIVER SARY!}>>. So, the answer is <<{\tt WILLIAM FRIEDMAN 125 ANNIVERSARY!}>>.

The problem was devoted to the anniversary of a mathematician and cryptographer William Friedman known as <<The Father of American Cryptology>> (September 24, 1891~--- November 12, 1969).

This problem was solved completely by 68 participants from all categories. They all acted in the same way. We could mention the best solutions of school students Alexander Grebennikov and Alexander Dorokhin (Presidential Physics and Mathematics Lyceum 239, Saint Petersburg), Arina Prostakova (Gymnasium 94, Yekaterinburg).

\subsection{Problem ``Get an access''}

{\bf Solution.} Here we want to describe a very compact and full solution by Vladimir Schavelev (Presidential Physics and Mathematics Lyceum 239, Saint Petersburg).

Let us prove that the sum can be any even number from 44 up to 76. It is obvious that the whole sum is even number since sums of any two numbers is even number (4,6 or 8). Let us consider the set where the sum takes its minimal possible value. Due to condition of the problem we have two numbers $a$ and $b$ with the sum equal to 8, so split all the numbers in pairs, such that one of these pairs is $a$, $b$. So, the minimal sum of such pairs is $8+9*4=44$. In the same way we obtain that the maximal possible sum is $4+9*8=76$.

If we fill cells with two 2 and the rest of the numbers are 4, we obtain the sum $4*18+4=76$, and this filling satisfies the condition. If we substitute any ``4'' by ``2'' we obtain the sum 74. We can continue this process, obtaining all possible variants of the sum, up to minimal, that is 44.

This problem was solved by 9 school students in the first round.

\subsection{Problem ``Find the key''}

{\bf Solution.} The answer for this problem is any set of positive integers $a$, $b$, $c$, $d$, $e$, $f$, $g$ such that the following relation holds:

\begin{center}
$a^3+b^3+c^3+d^3+e^3+f^3+g^3=2016^{2017}.$
\end{center}

For example, such a set can be found in the following way:
$$a^3+b^3+c^3+d^3+e^3+f^3+g^3=2016^{2017}=2016\cdot2016^{2016}=2016\cdot(2016^{672})^3.$$

Let us divide both sides on $2016^{2016}$ and assume that there exist $a'$, $b'$, $c'$, $d'$, $e'$, $f'$, $g'$ such that $a'^3+b'^3+c'^3+d'^3+e'^3+f'^3+g'^3=2016$. Then we can find easily these numbers, for instance, one of such sets is 3, 4, 5, 6, 7, 8, 9.

Then the original solution has the form $x=x'\cdot2016^{672}$. So, we have
$$(3\cdot2016^{672})^3+(4\cdot2016^{672})^3+(5\cdot2016^{672})^3+(6\cdot2016^{672})^3+(7\cdot2016^{672})^3+$$
$$(8\cdot2016^{672})^3+(9\cdot2016^{672})^3=2016^{2017}.$$

There were great solutions from Alexandr Grebennikov (Presidential Physics and Mathematics Lyceum 239, Saint Petersburg), Vadzim Marchuk, Anna	Gusakova, and Yuliya	Yarashenia team (Research Institute for Applied Problems of Mathematics and Informatics, Institut of Mathematics, Belarusian State University), that contains a lot of such keys and even the solutions by Alexey Ripinen, Oleg Smirnov, and Peter Razumovsky team (Saratov State University), Henning Seidler and Katja Stumpp team (Technical University of Berlin) that describes all possible keys. These solutions were awarded by additional scores.

\subsection{Problem ``Labyrinth''}

{\bf Solution.} Since there is a labyrinth, one can conjecture that the ciphertext is hidden in the right path from a mouse to cheese. Such a path is unique and contains the following string:
{\tt ONFIWQHWJJFLHZAOAXWESPPNGRCTPXGJXFWUDTOXYMCWJKML}.

The first hint given tells us that the secret message begins with <<{\tt ONE...}>>. By comparing the first three letters <<{\tt ONF}>> of the ciphertext with <<{\tt ONE}>> one can suppose that a polyalphabetic cipher is used. The hint about turns gives us an idea that each simple cipher is a substitution Ceasar cipher with some shift, where each turn in the labyrinth increase shift in the alphabet (Table\;\ref{Sol4}).

\begin{table}[ht]
\centering\footnotesize
\caption{{\bf Shifts}}
\label{Sol4}
\medskip
\renewcommand{\arraystretch}{1.2}
\renewcommand{\tabcolsep}{1.05mm}
\begin{tabular}{|cp{0.025\linewidth}p{0.025\linewidth}p{0.025\linewidth}p{0.025\linewidth}p{0.025\linewidth}p{0.025\linewidth}p{0.025\linewidth}p{0.025\linewidth}p{0.025\linewidth}p{0.025\linewidth}p{0.025\linewidth}p{0.025\linewidth}p{0.025\linewidth}p{0.025\linewidth}p{0.025\linewidth}p{0.025\linewidth}p{0.025\linewidth}p{0.025\linewidth}p{0.025\linewidth}p{0.025\linewidth}p{0.025\linewidth}p{0.025\linewidth}p{0.025\linewidth}p{0.025\linewidth}|}
\hline
Path & {\tt O} & {\tt N} & {\tt F} & {\tt I} & {\tt W} & {\tt Q} & {\tt H} & {\tt W} & {\tt J} & {\tt J} & {\tt F} & {\tt L} & {\tt H} & {\tt Z} & {\tt A} & {\tt O} & {\tt A} & {\tt X} & {\tt W} & {\tt E} & {\tt S} & {\tt P} & {\tt P} & {\tt N} \\
Shift &  0 &  0 & 1 & 1 & 2 & 3  & 4 & 5 & 5 & 6 & 7 & 7 & 7 & 8 & 8 & 9 & 9 & 9 & 10 & 11 & 11 & 11 & 12 & 13 \\
Message & {\tt O} & {\tt N} & {\tt E} & {\tt H} & {\tt U} & {\tt N} & {\tt D} & {\tt R} & {\tt E} & {\tt D} & {\tt Y} & {\tt E} & {\tt A} & {\tt R} & {\tt S} & {\tt F} & {\tt R} & {\tt O} & {\tt M} & {\tt T} & {\tt H} & {\tt E} & {\tt D} & {\tt A}\\
\hline
\hline
Path & {\tt G} & {\tt R} & {\tt C} & {\tt T} & {\tt P} & {\tt X} & {\tt G} & {\tt J} & {\tt X} & {\tt F} & {\tt W} & {\tt U} & {\tt D} & {\tt T} & {\tt O} & {\tt X} & {\tt Y} & {\tt M} & {\tt C} & {\tt W} & {\tt J} & {\tt K} & {\tt M} & {\tt L} \\
Shift & 13 & 13 & 14 & 14 & 14 & 15 & 15 & 16 & 16 & 17 & 17 & 18 & 18 & 19 & 20 & 20 & 20 & 20 & 21 & 22 & 22 & 23 & 24 & 24\\
Message & {\tt T} & {\tt E} & {\tt O} & {\tt F} & {\tt B} & {\tt I} & {\tt R} & {\tt T} & {\tt H} & {\tt O} & {\tt F} & {\tt C} & {\tt L} & {\tt A} & {\tt U} & {\tt D} & {\tt E} & {\tt S} & {\tt H} & {\tt A} & {\tt N} & {\tt N} & {\tt O} & {\tt N}\\
\hline
\end{tabular}
\end{table}

Thus, the secrete message can be read as:
\begin{center}
<<One hundred years from the date of birth of Claude Shannon>>.
\end{center}

It was devoted to the anniversary of an outstanding mathematician, electrical engineer, and cryptographer Claude Elwood Shannon (April 30, 1916 --- February 24, 2001).

The problem was completely solved by 46 participants of all categories in both rounds. The most beautiful solutions were presented by Arina Prostakova (Gymnasium 94, Yekaterinburg), Maria Tarabarina (Lomonosov Moscow State University), Dragos Alin Rotaru, Marco Martinoli, and Tim Wood team (University of Bristol, United Kingdom), Nguyen Duc, Bui Minh Tien Dat, and Quan Doan team (University of Informa\-tion Technology, Vietnam), Henning Seidler and Katja Stumpp team (Technical University of Berlin).

\subsection{Problem ``System of equations''}

{\bf Solution.} One can notice these equations can be grouped under the three subsystems:
\begin{center}$
 1) \begin{cases}
   x_1\oplus x_2\oplus x_3\oplus x_5\oplus x_6=0, \\
   x_1x_3\oplus x_2x_4=x_5-x_6, \\
   (x_2+x_3+x_4)^2=2(x_1+x_5+x_6), \\
   x_2x_3x_4\oplus x_1x_5x_6=0; \\
 \end{cases}
$
$
 2) \begin{cases}
   x_6\oplus x_7\oplus x_8\oplus x_{10}\oplus x_{11}=0, \\
   x_6x_8\oplus x_7x_9=x_{10}-x_{11}, \\
   (x_7+x_8+x_9)^2=2(x_6+x_{10}+x_{11}), \\
   x_7x_8x_9\oplus x_6x_{10}x_{11}=0; \\
 \end{cases}
$

$
 3) \begin{cases}
   x_{11}\oplus x_{12}\oplus x_{13}\oplus x_{15}\oplus x_{16}=0, \\
    x_{11}x_{13}\oplus x_{12}x_{14}=x_{15}-x_{16}, \\
   (x_{12}+x_{13}+x_{14})^2=2(x_{11}+x_{15}+x_{16}), \\
   x_{11}x_{12}x_{13}\oplus x_{14}x_{15}x_{16}=0. \\
 \end{cases}
$
\end{center}
Note that the first and the second subsystems have a common variable $x_6$, just like the second and the third ones both involve $x_{11}$.

The first two are the subsystems having the template
\begin{equation*}
 \begin{cases}
   y_1\oplus y_2\oplus y_3\oplus y_5\oplus y_6=0, \\
   y_1y_3\oplus y_2y_4=y_5-y_6, \\
   (y_2+y_3+y_4)^2=2(y_1+y_5+y_6), \\
   y_2y_3y_4\oplus y_1y_5y_6=0,
 \end{cases}
\end{equation*}

but the third subsystem has the following one
\begin{equation*}
 \begin{cases}
   y_1\oplus y_2\oplus y_3\oplus y_5\oplus y_6=0, \\
   y_1y_3\oplus y_2y_4=y_5-y_6, \\
   (y_2+y_3+y_4)^2=2(y_1+y_5+y_6), \\
   y_1y_2y_3\oplus y_4y_5y_6=0.
 \end{cases}
\end{equation*}

In both of them variables are $\left(y_1,y_2,y_3,y_4,y_5,y_6\right)={\bf y}$, where $y_i\in\{0,1\}$, $i=~1,2,\ldots,6$.

Obviously, one of the solutions of both templates is equal to ${\bf y}=(0,0,0,0,0,0)$; we denote it by ${\bf y}^1$.

Let us consider the first template and find all its solutions.

In the case $\underline{y_5=y_6}=y\in\{0,1\}$ we have
\begin{equation*}
 \begin{cases}
   y_1\oplus y_2\oplus y_3=0, \\
   y_1y_3\oplus y_2y_4=0, \\
   (y_2+y_3+y_4)^2=2(y_1+2y), \\
   y_2y_3y_4\oplus y_1y=0. \\
 \end{cases}
\end{equation*}

Evidently, the third equation holds if
\begin{center}
a) $y_2+y_3+y_4=0$ and $y_1+2y=0$;
\ \ or\ \
b) $y_2+y_3+y_4=2$ and $y_1+2y=2$.
\end{center}

From a) we obtain ${\bf y}^1$. Using the case b) and the equation $y_1\oplus y_2\oplus y_3=0$ we receive $y=1$, $y_1=0$, $y_2\oplus y_3=0$ (i.\,e. $y_2=y_3=y'\in\{0,1\}$) and $2y'+y_4=2$. Thus, we have ${\bf y}=(0,1,1,0,1,1)$, which evidently satisfies the other equations of the template. So, the second solution is $(0,1,1,0,1,1)$; we denote it by ${\bf y}^2$.

In the case $\underline{y_5\ne y_6}$ (i.\,e. $y_5=1$, $y_6=0$ since it must hold $y_5\geqslant y_6$ by virtue of the second equation) we have
\begin{equation*}
 \begin{cases}
   y_1\oplus y_2\oplus y_3\oplus1=0, \\
   y_1y_3\oplus y_2y_4=1, \\
   (y_2+y_3+y_4)^2=2(y_1+1), \\
   y_2y_3y_4=0. \\
 \end{cases}
\end{equation*}

Again, the third equation holds if
\begin{center}
a) $y_2+y_3+y_4=0$ and $y_1+1=0$;
\ \ or\ \
b) $y_2+y_3+y_4=2$ and $y_1+1=2$.
\end{center}

The case a) is impossible for binary $y_1$. Using the case b) and the equation $y_1\oplus y_2\oplus y_3=1$ we receive $y_1=1$, $y_2\oplus y_3=0$ (i.\,e. $y_2=y_3=y''\in\{0,1\}$) and $2y''+y_4=2$. In this way we have ${\bf y}=(1,1,1,0,1,0)$, which evidently satisfies the other equations of the template. So, the third solution is $(1,1,1,0,1,0)$; we denote it by ${\bf y}^3$.

Let us consider the second template. It only differs from the first template in the fourth equation. But at the same time, this equation has not been used while obtaining the solutions besides checking. So, it is enough to check whether these solutions satisfy the equation $y_1y_2y_3\oplus y_4y_5y_6=0$. Obviously, ${\bf y}^1$ and ${\bf y}^2$ are suitable, but ${\bf y}^3$ does not satisfy it.

Thus, after considering the links between corresponding subsystems that involve common variables, we get that all solutions of the initial system are the following:
\begin{center}
\begin{tabular}{lll}
1. $(00000{\bf0}0000{\bf0}00000)$; & 3. $(01101{\bf1}1101{\bf0}00000)$; & 5. $(11101{\bf0}0000{\bf0}00000)$;\\
2. $(00000{\bf0}0000{\bf0}11011)$; & 4. $(01101{\bf1}1101{\bf0}11011)$; & 6. $(11101{\bf0}0000{\bf0}11011)$.\\
\end{tabular}
\end{center}

The problem was completely solved by 26 participants, 12 of whom used computer calculations. Many participants have noticed it is possible to solve the system separately by taking into account the structure of the whole system. But at the same time, some of them have not considered the difference between the templates, in such cases extra solutions have been included in the problem's answer. The right solutions with good explanation were made by Alexandr Grebennikov (Presidential
Physics and Mathematics Lyceum 239, Saint Petersburg), Dmitry Morozov (Novosibirsk State University), Anna Gusakova (Institute of Mathematics of National Academy of Sciences of Belarus).

\subsection{Problem ``Biometric pin-code''}

{\bf Solution.} First we compute $s'=b\oplus c$, as described in the algoritm, and obtain $$s'=(1000\:0111\:1101\:0000).$$ Note that, since a new biometric image $b$ of Alice can be different from the image that was taken during creation of her biometrically encrypted key (by not more than 10--20\%), $s'$ can also be different from Hadamard code codeword $s$ corresponding to real Alice's key $k$, but by 10--20\% at most.

So, in order to find Alice's key, we need to find a codeword of Hadamard code of length 16, which differs from $s'$ in not more than $20\%$ of bits. Since all codewords of Hadamard code differ from each other in at least 50\% of bits, such a codeword is unique (in case it exists). Thus, we can simply search for the codeword closest to $s'$.

Since Hadamard code of length 16 has $2^5=32$ codewords, we can easily (with or without use of computer) find the closest codeword: it is $s=(0000\:1111\:1111\:0000)$ and it corresponds to the key $k=(11000)$. We see that $s$ and $s'$ differ in only 3 bits, which is $18.75\%$ of all 16 bits. So the key $k$ that we have found fits all conditions. Therefore, it is the real key of Alice.

Unfortunately, there were no complete solutions of the problem by school students.

\subsection{Problem ``Quadratic functions''}

{\bf Solution.} Here we would like to describe the solution by George Beloshapko, Stepan Gatilov, and Anna Taranenko team (Novosibirsk State University, Ledas, Sobolev Institute of Mathema\-tics) as the simplest.

Let us consider the sequences $\{u_i\}$ and $\{v_i\}$ generated by the functions $f_u(x_1, \ldots, x_n) = x_1 x_2$ and $f_v(x_1, \ldots, x_n) = x_2 x_n \oplus x_1 \oplus x_n$ respectively with the initial value
$$
	\underbrace{1 \ldots 10}_n.
$$

Let us describe the first $n^2 - n + 1$ elements of the sequences:
\begin{eqnarray*}
	\{u_i\} & = & \underbrace{1 \ldots 1}_{n - 1}\underbrace{0}_{1} \underbrace{1 \ldots 1}_{n - 2}\underbrace{00}_{2} \ldots \underbrace{11}_{2}\underbrace{0 \ldots 0}_{n - 2}\underbrace{1}_{1}\underbrace{0 \ldots 0}_{n - 1} 0... \\	
	\{v_i\} & = & \underbrace{1 \ldots 1}_{n - 1}\underbrace{0}_{1} \underbrace{1 \ldots 1}_{n - 2}\underbrace{00}_{2} \ldots \underbrace{11}_{2}\underbrace{0 \ldots 0}_{n - 2}\underbrace{1}_{1}\underbrace{0 \ldots 0}_{n - 1} 1... \\
\end{eqnarray*}

The first $n^2 - n$ elements of the sequences are the same, but the next elements differ.
So, it is impossible to uniquely reconstruct the sequence by the segment of length $cn$, where $c$ is a constant.

There were no complete solutions of the problem in the first round.
In the second round, in addition to the given solution, the problem was completely solved by  Alexey Ripinen, Oleg Smirnov, and Peter Razumovsky team (Saratov	Saratov State University), Maxim Plushkin, Ivan Lozinskiy, and Alexey Solovev team (Lomonosov Moscow State University), Alexey Udovenko (University of Luxembourg).

\subsection{Problem ``Biometric key''}

{\bf Solution.} To solve the problem, we compute values $s'$ for both persons $X$ and $Y$ and check if any of them is close enough to some Hadamard code codeword, corresponding to the key with odd number of 1's. Here ``close enough'' means difference in not more than 20\% bits, since two biometric images of the same person can not be more different from each other.

Let us denote these values $s^X$ and $s^Y$ for person $X$ and $Y$ respectively. Now, with the use of a computer program, we can go through all 256 codewords of Hadamard code of length 128, and find those which are closest to $s^X$ and $s^Y$.

For person $Y$, the closest Hadamard codeword differs from $s^Y$ by 49 bits (around 38.28\%). This means that there exists no key $k$, using which person $Y$ could make himself biometrically encrypted key $c$.

For person $X$, the closest Hadamard codeword is obtained from $(11011010)$ and it differs from $s^X$ by 25 bits (around 19.53\%). This means that person $X$ is Alice and her 8-bit key is $k=(11011010)$ (it also has an odd number of 1's, which solidifies our confidence).

Note that there are many interesting approaches for combining biometrics with cryptography (for example, see \cite{11-Rathgeb}).

The problem was completely solved by Irina Slonkina (Novosibirsk State University of Economics and Management) and George Beloshapko (Novosibirsk State University) in the first round, and by 10 teams in the second round, Evgeniya Ishchukova, Ekaterina Maro, and Dmitry Alekseev team (Southern Federal University, Taganrog) was among them.

\subsection{Problem ``Secret sharing''}

{\bf Solution.} Without Sergey, Alena and Boris know the following vector:
$$
	P' = P \oplus \bigoplus_{i = 1}^{32}{c_i^s v_i^s}.
$$
So, they do not know any information about $P$ if and only if the dimension of the linear span $\langle v_1^s, \ldots, v_{32}^s \rangle$ is equal to $32$, i.\,e. $v_1^s, \ldots, v_{32}^s$ form basis of $\F{32}$. Otherwise,  $\langle v_1^s, \ldots, v_{32}^s \rangle \subset \F{32}$ and $P \in P' \oplus \langle v_1^s, \ldots, v_{32}^s \rangle$, it means that there are not more than $2^{\text{rk} \langle v_1^s, \ldots, v_{32}^s \rangle}$ ways for Alena and Boris to get $P$.

Let $V$ be a $32 \times 32$ matrix with columns $v_1^s, \ldots, v_{32}^s$. Note that $\text{rk} V = \text{rk} \langle v_1^s, \ldots, v_{32}^s \rangle$. Since all bits of $v_1^s, \ldots, v_{32}^s$ are randomly generated (and independent of each other), all elements of $V$ are randomly generated too.

Hence, the probability $p_1$ that Alena and Boris can not get any information about $P$ is equal to $N_1 / 2^{32 \cdot 32}$, where $N_1$ is the number of matrices $V$ of rank $32$:
$$
	p_1 = \frac{(2^{32} - 2^0) \cdot (2^{32} - 2^1) \cdot \ldots \cdot (2^{32} - 2^{31})}{2^{32 \cdot 32}} \approx 0,288788.
$$

In order for Alena and Boris to get a guaranteed access to online banking without Sergey using not more than 23 attempts, it should hold $2^{\text{rk}\langle v_1^s, \ldots, v_{32}^s \rangle} \leq 23$, i.\,e. $\text{rk}\langle v_1^s, \ldots, v_{32}^s \rangle
\leq 4$.
So, the probability $p_2$ of that is equal to $N_2 / 2^{32 \cdot 32}$, where $N_2$ is the number of matrices $V$ of rank not more than $4$.

Note that the number of $n\times n$ matrices of rank $k$ is equal to
$$
	R_n^k = \frac{(2^{n} - 2^0)^2 \cdot (2^{n} - 2^1)^2 \cdot \ldots \cdot (2^{n} - 2^{k - 1})^2}
	{(2^{k} - 2^0) \cdot (2^{k} - 2^1) \cdot \ldots \cdot (2^{k} - 2^{k - 1})}.
$$
Therefore,
$$
	p_2 = \frac{R_{32}^0 + R_{32}^1 + R_{32}^2 + R_{32}^3 + R_{32}^4}{2^{32 \cdot 32}} \approx 1.625 \cdot 2^{-783}.
$$

In the first round, the problem was completely solved by Alexey Udovenko (University of Luxembourg) and Igor Fedorov (Novosibirsk State University); almost completely solved by Mohammadjavad Hajialikhani (Sharif University of Technology, Iran), Pavel Hvoryh (Omsk State Technical University), George Beloshapko (Novosibirsk State University) and Ekaterina Kulikova (Munich, Germany).
In the second round (in addition to the first round) the problem was completely solved by Aliaksei Ivanin, Oleg Volodko, and Konstantin Pavlov team (Belarusian State University).

\subsection{Problem ``Protocol''}

{\bf Solution.} Here we would like to describe the solution proposed by Alexey Udovenko (University of Luxembourg), that is similar to the author's one, but is more elegant and compact.

The session key is equal to
$$K_{a,b}=g^{R_aR_b} =g^{(X_a-\alpha_a)(X_b-\alpha_b)}=g^{X_aX_b-\alpha_aX_b-\alpha_bX_a+\alpha_a\alpha_b}\!\!\pmod{p}.$$

Evgeniy observes $X_a$ and $X_b$ and he also knows $g$, $P_a =
g^{\alpha_a}\!\!\mod{p}$ and $P_b = g^{\alpha_b}\!\!\mod{p}$. From $g, P_a, P_b$ and one
exposured key he can compute
$$ s = g^{\alpha_a\alpha_b} =
K_{a,b}/(g^{X_aX_b}P_a^{-X_b}P_b^{-X_a})\!\!\pmod{p}.$$
Then for new
sessions he can intercept new $X_a$ and $X_b$ and easily compute new
$$K'_{a,b} = g^{X_aX_b}P_a^{-X_b}P_b^{-X_a}s\!\!\mod{p}.$$

It worth noting that recovering of the next keys would be impossible if the protocol has a property of so-called {\it forward secrecy}. More details can be found in \cite{92-Diffie}.

This problem was completely solved by 7 participants in the first round, Robert Spencer (University of Cape Town, South Africa) was among them, and by 15 teams in the second round, Roman Lebedev, Ilia Koriakin, and Vlad Kuzin team (Novosibirsk State University) was among them.
All solutions proposed were made in the similar way. Also there were
solutions with reduced score, containing some inexactness, but they
were still very close to complete ones.

\subsection{Problem ``Zerosum at AES''}

{\bf Solution.} Let us describe one of the simplest approach to get the solution proposed and implemented by Alexey Udovenko (University of Luxembourg).

Denote by $Y_i = AES_0(X_i) \oplus X_i$. The equation can be rewritten:
$$
	\bigoplus_{i = 1}^{128} (AES_0(X_i) \oplus X_i) = \bigoplus_{i = 1}^{128} Y_i = 0.
$$
First we encrypt random $256$ plaintexts and obtain $Y_1, \ldots, Y_{256}$.
Then the problem is to find distinct indices $i_1, \ldots, i_{128}$ such that $Y_{i_1} \oplus \ldots \oplus Y_{i_{128}} = 0$.

Let $M$ be the $128 \times 256$ binary matrix with columns $Y_1, \ldots, Y_{256}$. Let us consider the solutions of the linear equation $M z = 0$: for any solution $z$ it holds
$$
	\bigoplus_{i: z_i = 1} Y_i = 0.
$$
Moreover, the Hamming weight of a random solution vector $z$ will be close to $128$ and with high probability equal to $128$. Next, we try to find a solution vector of weight $128$.

All information about the block cipher AES can be found in the book \cite{02-DaeRij} of AES authors J.\;Daemen and V.\;Rijmen.

The problem was completely solved (proposed an algorithm and the solution) by three teams: Maxim Plushkin, Ivan Lozinskiy, and Alexey Solovev team (Lomonosov Moscow State University), Alexey Udovenko (University of Luxembourg), George Beloshapko, Stepan Gatilov, and Anna Taranenko team (Novosibirsk State University, Ledas, Sobolev Institute of Mathematics).

\subsection{Problem ``Latin square''}

{\bf Solution.}
A $n\times n$ latin square  can be given as the set of its columns permutation $\sigma_i$, $i=1,\ldots,n$.
Then, the answer $t_3$ on the request $abcd$ can be calculated as~$\sigma_d\sigma_c\sigma_b(a)$ (here $\sigma=\sigma_d\sigma_c\sigma_b$ denotes the permutations composition).
Choosing all possible $a\neq b$, one is able to reconstruct~$\sigma$. It is needed 9 request--answer pairs.
The inverse permutations can also be found, $\sigma^{-1}=\sigma_b^{-1}\sigma_c^{-1}\sigma_d^{-1}$.

Moreover, for any distinct $i,j\in\{0,1,\ldots,9\}$ the following permutation is recovered using
18 request--answer pairs:
$$
\sigma_j^{-1}\sigma_i=
(\sigma_j^{-1}\sigma_c^{-1}\sigma_d^{-1})
(\sigma_d\sigma_c\sigma_i).
$$
Here $c,d$ are arbitrary numbers from the set $\{0,1,\ldots,9\}\setminus\{i,j\}$.

If one knows $\sigma_j^{-1}\sigma_i$, then he also knows 10 preimage--image pairs $x,y$:
$y=\sigma_j^{-1}\sigma_i(x)$.
Each of such a pair means an equality~$\sigma_i(x)=\sigma_j(y)$, from which
$\sigma_j$ can be expressed by~$\sigma_i$.

For example, let us express $\sigma_j$, $j=1,2,\ldots,9$, by~$\sigma_0$.
To do this we need $18\cdot 9=162$ pairs of request--answer.
Then we will check all possible variants of $\sigma_0$ (there are $10!$ such variants).
For each variant $\hat{\sigma}_0$ we find~$\hat{\sigma}_j$, build the corresponding latin square~$\hat{L}$ and check whether answers on $\hat{L}$ is equal to answers on the secret latin square.
Using~$162\gg\log_{10}10!$ answers Alice's secret key will be uniquely recovered
with high probability.

The answer is the following $10\times 10$ latin square.

\begin{center}
\begin{tabular}{|c|c|c|c|c|c|c|c|c|c|}
	\hline
	3 & 9 & 2 & 1 & 7 & 8 & 5 & 0 & 6 & 4 \\
	\hline
	4 & 3 & 7 & 9 & 1 & 5 & 2 & 6 & 8 & 0 \\
	\hline
	1 & 8 & 6 & 7 & 0 & 9 & 4 & 3 & 5 & 2 \\
	\hline
	2 & 6 & 5 & 3 & 9 & 7 & 1 & 4 & 0 & 8 \\
	\hline
	6 & 0 & 9 & 4 & 5 & 1 & 8 & 2 & 3 & 7 \\
	\hline
	0 & 7 & 3 & 6 & 2 & 4 & 9 & 8 & 1 & 5 \\
	\hline
	8 & 4 & 1 & 0 & 3 & 2 & 6 & 5 & 7 & 9 \\
	\hline
	5 & 1 & 4 & 8 & 6 & 0 & 7 & 9 & 2 & 3 \\
	\hline
	7 & 2 & 8 & 5 & 4 & 3 & 0 & 1 & 9 & 6 \\
	\hline
	9 & 5 & 0 & 2 & 8 & 6 & 3 & 7 & 4 & 1 \\
	\hline
\end{tabular}
\end{center}

The problem was completely solved by $26$ teams. The best solution was proposed by Sergey Titov (Ural State University of Railway Transport, Yekaterinburg).

\subsection{Problem ``\textsl{nsucoin}''}

{\bf Solution.} The payment method used is an example of blockchain based money working on the proof-of-work principal. The first such a system \textsl{bitcoin} was proposed in \cite{09-Nakamoto}.

 To solve this problem one need to restore a history of transactions that leads to the block on height 2 from condition of the problem:
\begin{center}
{\tt height:2;prevHash:0000593b;ctxHash:8fef76cb;nonce:17052}
\end{center}
Given such a history of transactions one can find how many flowers of different kinds Alice has at the end of trading.

\smallskip
 Solution plan consists of the following steps:

{\bf1}. One need to find each user's private key (using private keys one is able to make a sign of each user and as a result generate transactions).

{\bf2}. One need to find all special transactions that give to each user 10 coins.

{\bf3}. One need to find all possible blocks on height 0.

{\bf4}. Looking through all possible transactions, one need to find blocks on height 1 for each block on height 0 so that hash of each block found will be equal to value of {\texttt prevHash} field of the  given block, i.\,e. {\texttt 0000593b}.

{\bf5}. It is known that there is at least one block among the blocks found on the previous step such that there exist transactions such that hash of these transactions concatenation is equal to {\texttt 8fef76cb}. One need to find these transactions.

{\bf6}. Thus, all transactions history leading to the given block on height 2 is found. It remains to track the movement of flowers and give an answer.

While searching blocks we need to remember that {\tt nonce} does not exceed 40000. Despite the fact that there are 24 transpositions of special transactions, only 6 of them can be verified by a block. Also, it is useful to remember that each participant can sell only 5 flowers, each of them costs 2 coins, and a participant can not buy anything if he does not have coins.

It is easy to factor a given small module of RSA:
$$n = p\cdot q = 2250339337\cdot 4044301367 = 9101050456842973679,$$
$$\phi(n) = (p-1)(q-1) = 9101050450548332976.$$
Then, one can find private keys as inverse numbers to public keys module $\phi(n)$ (Table\;\ref{Tab-keys}).

\begin{table}[ht]
\centering\footnotesize
\caption{{\bf Public and private keys}}
\medskip
\label{Tab-keys}
  \begin{tabular}{|c|c|c|}
    \hline
    User & pubKey & privKey \\
    \hline
    \hline
    {\text Alice} & 11 & 2482104668331363539 \\
    \hline
    {\text Bob} & 17 & 3747491361990490049\\
    \hline
    {\text Caroline} & 199 & 9009582606824229127\\
    \hline
    {\text Daniel} & 5 & 7280840360438666381\\
    \hline
  \end{tabular}
\end{table}

Despite all the limitations of searching, it turns out to be quite time-consuming. There were laid some hints in the condition of the problem allowing one to get a solution.
This two hints were published on the Olympiad website:

\underline{18 November}. A tip to reduce exhaustive search: The transaction from the example has been verified by the block with height 1.

\underline{20 November}. All hashes of transactions that are verified by the block from the example correspond to hashes of transactions verified by the block with height 1 into the sought-for blockchain.

The first participant who found out the right answer was Alexey Udovenko (University of Luxembourg). He based only on the first hint and guessed that all four transactions hashes from the example are contained in the block with height 1. By the other words, he guessed the second hint and found the answer we had conceived.

In fact, we knew only one answer but we hoped that someone was able to find another one, and it happened! There were two teams that found two answers based only on the first hint: George Beloshapko, Stepan Gatilov, and Anna Taranenko team (Novosibirsk State University, Ledas, Institute of Mathematics) and the team of Alexey Ripinen, Oleg Smirnov, and Peter Razumovsky team (Saratov State University).

 The final flowers distribution and transactions history  of two solutions are presented in Tables\;\ref{Tab-flowers},\,\ref{Tab-trans}.

 There were 9 teams received scores for this problem, 2 teams found two answers, 1 found an answer based only on the first hint, 2 found answer based on both hints, 3 teams found blocks with height 0 and 1 but did not found transactions verified by block with height 2, and 1 team found only a block with height 0. Probably, someone were not able to find answer due to the following fact:
each transaction to be verified can be based not only on already verified transactions but also on transactions based on verified transactions. It is transactions of the second type that are in blocks. Also, the difficulties for the decisive could create transactions based on two other transactions. There is one such transaction verified by the block with height 2 in the first answer and four such transactions in the second answer.

\begin{table}[ht]
\centering\footnotesize
\caption{{\bf Flowers distribution of two solutions}}
\label{Tab-flowers}
\medskip
\setlength{\tabcolsep}{0.1cm}
  \begin{tabular}{|c||c|c|c|c|c|c|c|c|}
  \hline
    \ & \multicolumn{2}{c|}{camomiles} & \multicolumn{2}{c|}{tulips} &
    \multicolumn{2}{c|}{gerberas}  & \multicolumn{2}{c|}{roses}
    \\
    \hline
    & 1st sol. & 2d sol. & 1st sol. & 2d sol.& 1st sol. & 2d sol.& 1st sol. & 2d sol.\\
    \hline
\hline
   Alice &  1 & 1 & 2 & 3 &3 & 0 & 1 & 0 \\
    \hline
   Bob & 4 & 2& 0 & 2& 2 & 2 & 0 & 1 \\
    \hline
   Caroline & 0 & 2& 0 & 0& 0 & 1 & 1 &0 \\
    \hline
   Daniel & 0 & 0 & 3 & 0& 0 &2 & 3  & 4\\
  \hline
  \end{tabular}
\end{table}

\begin{table}[p]
\centering\footnotesize
	\rotatebox{90}{
		\begin{minipage}{\textheight}
\caption{{\bf Transactions history of two solutions}}
\label{Tab-trans}
\medskip
\renewcommand{\arraystretch}{0.8}
\renewcommand{\tabcolsep}{0.12cm}
\begin{tabular}{l}
{\bf First solution}:\\
Tx1: {\tt\footnotesize txHash:2c8993af;inputTx:;value1:10;pubKey1:199;sign1: }\\
Tx2: {\tt\footnotesize txHash:1a497b59;inputTx:;value1:10;pubKey1:11;sign1: }\\
Tx3: {\tt\footnotesize  txHash:4d272154;inputTx:;value1:10;pubKey1:17;sign1:}\\
Tx4: {\tt\footnotesize  txHash:05722480;inputTx:;value1:10;pubKey1:5;sign1:}\\
Block 1: {\tt\footnotesize height:0;prevHash:00000000;ctxHash:bde430dd;nonce:21095} {\footnotesize with hash} {\tt\footnotesize 00003cc0}\\
Tx5: {\tt\footnotesize txHash:98e93fd5;inputTx:1a497b59;value1:4;pubKey1:17;sign1:689e297682a9e6f0;value2:6;pubKey2:11;sign2:fec9245898b829c}\\
Tx6: {\tt\footnotesize  txHash:c16d8b22;inputTx:98e93fd5;value1:6;pubKey1:199;sign1:611cddad83c8e352;value2:0;pubKey2:11;sign2:6ab37bb9f5f4249d}\\
Tx7: {\tt\footnotesize txHash:b782c145;inputTx:4d272154;value1:4;pubKey1:199;sign1:5f50adf4a51899e5;value2:6;pubKey2:17;sign2:78a4f7d6d0d578d7}\\
Tx8: {\tt\footnotesize  txHash:e1e2c554;inputTx:05722480;value1:6;pubKey1:17;sign1:b91ab453e4bcb53;value2:4;pubKey2:5;sign2:3803907748416c12}\\
Block 2: {\tt\footnotesize  height:1;prevHash:00003cc0;ctxHash:9f8333d4;nonce:21438} {\footnotesize with hash} {\tt\footnotesize 0000593b}\\
Tx9: {\tt\footnotesize txHash:641c33ac;inputTx:98e93fd5,e1e2c554;value1:8;pubKey1:11;sign1:110a7c6e1dfd937;value2:2;pubKey2:17;sign2:14ad50a36ef5540d}\\
Tx10: {\tt\footnotesize  txHash:3aff68cb;inputTx:641c33ac;value1:2;pubKey1:5;sign1:20281a4d62b20cb7;value2:6;pubKey2:11;sign2:7b662b128b1200bf}\\
Tx11: {\tt\footnotesize  txHash:3aff68cb;inputTx:641c33ac;value1:2;pubKey1:5;sign1:20281a4d62b20cb7;value2:6;pubKey2:11;sign2:7b662b128b1200bf}\\
Block 3: {\tt\footnotesize  height:2;prevHash:0000593b;ctxHash:8fef76cb;nonce:17052} {\footnotesize with hash} {\tt\footnotesize 000023d4}\\
\ \\
{\bf Second solution}:\\
Tx1: {\tt\footnotesize txHash:4d272154;inputTx:;value1:10;pubKey1:17;sign1: }\\
Tx2: {\tt\footnotesize txHash:1a497b59;inputTx:;value1:10;pubKey1:11;sign1: }\\
Tx3: {\tt\footnotesize txHash:05722480;inputTx:;value1:10;pubKey1:5;sign1:}\\
Tx4: {\tt\footnotesize txHash:2c8993af;inputTx:;value1:10;pubKey1:199;sign1:}\\
Block 1: {\tt\footnotesize height:0;prevHash:00000000;ctxHash:4ddc0244;nonce:20670} {\footnotesize with hash} {\tt\footnotesize 0000857a}\\
Tx5: {\tt\footnotesize txHash:txHash:5c3b1a45;inputTx:4d272154;value1:4;pubKey1:11;sign1:5866152e5bf782f1;value2:6;pubKey2:17;sign2:78a4f7d6d0d578d7}\\
Tx6: {\tt\footnotesize txHash:98e93fd5;inputTx:1a497b59;value1:4;pubKey1:17;sign1:689e297682a9e6f0;value2:6;pubKey2:11;sign2:fec9245898b829c}\\
Tx7: {\tt\footnotesize  txHash:f64f4e31;inputTx:98e93fd5,5c3b1a45;value1:2;pubKey1:17;sign1:6a9369e096c2cd05;value2:8;pubKey2:11;sign2:5021efb4fb05e703}\\
Tx8: {\tt\footnotesize txHash:ed107efb;inputTx:5c3b1a45;value1:2;pubKey1:199;sign1:7e12b526fd676d32;value2:4;pubKey2:17;sign2:4d9942d6d31e6392}\\
Block 2: {\tt\footnotesize height:1;prevHash:0000857a;ctxHash:00f229f7;nonce:19574} {\footnotesize with hash} {\tt\footnotesize 0000593b}\\
Tx9: {\tt\footnotesize txHash:bdddf6d7;inputTx:05722480;value1:4;pubKey1:199;sign1:6b33ced4b96ed36f;value2:6;pubKey2:5;sign2:3803907748416c12}\\
Tx10: {\tt\footnotesize txHash:211f6f39;inputTx:ed107efb,f64f4e31;value1:2;pubKey1:5;sign1:6b12e0f356e951ea;value2:4;pubKey2:17;sign2:88b48219f607775}\\
Tx11: {\tt\footnotesize txHash:944ac28f;inputTx:ed107efb,bdddf6d7;value1:4;pubKey1:11;sign1:395867768ee9f790;value2:2;pubKey2:199;sign2:555727e07c7e0c97}\\
Tx12: {\tt\footnotesize
txHash:e88eea0e;inputTx:211f6f39,98e93fd5;value1:2;pubKey1:199;sign1:76a91f809d7468b7;value2:6;pubKey2:17;sign2:1c6905596113f9a6}\\
Block 3: {\tt\footnotesize  height:2;prevHash:0000593b;ctxHash:8fef76cb;nonce:17052} {\footnotesize with hash} {\tt\footnotesize 000023d4}\\
\end{tabular}
\end{minipage}
}
\end{table}

Special thanks to the team of  George Beloshapko, Stepan Gatilov, and Anna Taranenko for illustrating the solutions:

\begin{center}
\includegraphics[height=6.4cm]{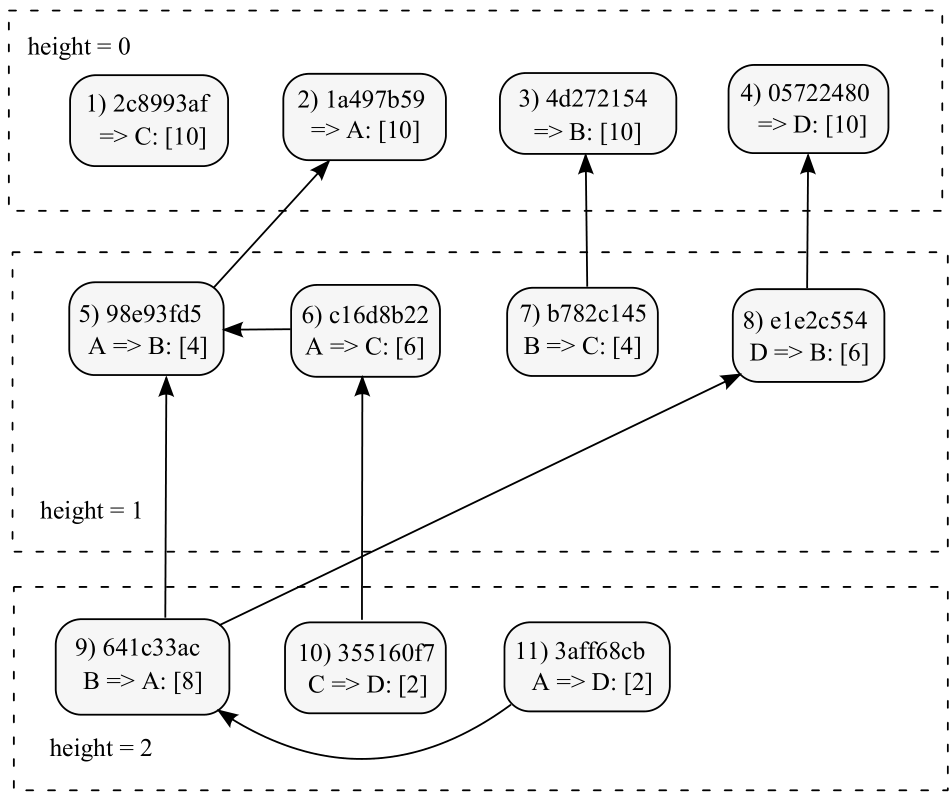}
\ \includegraphics[height=6.4cm]{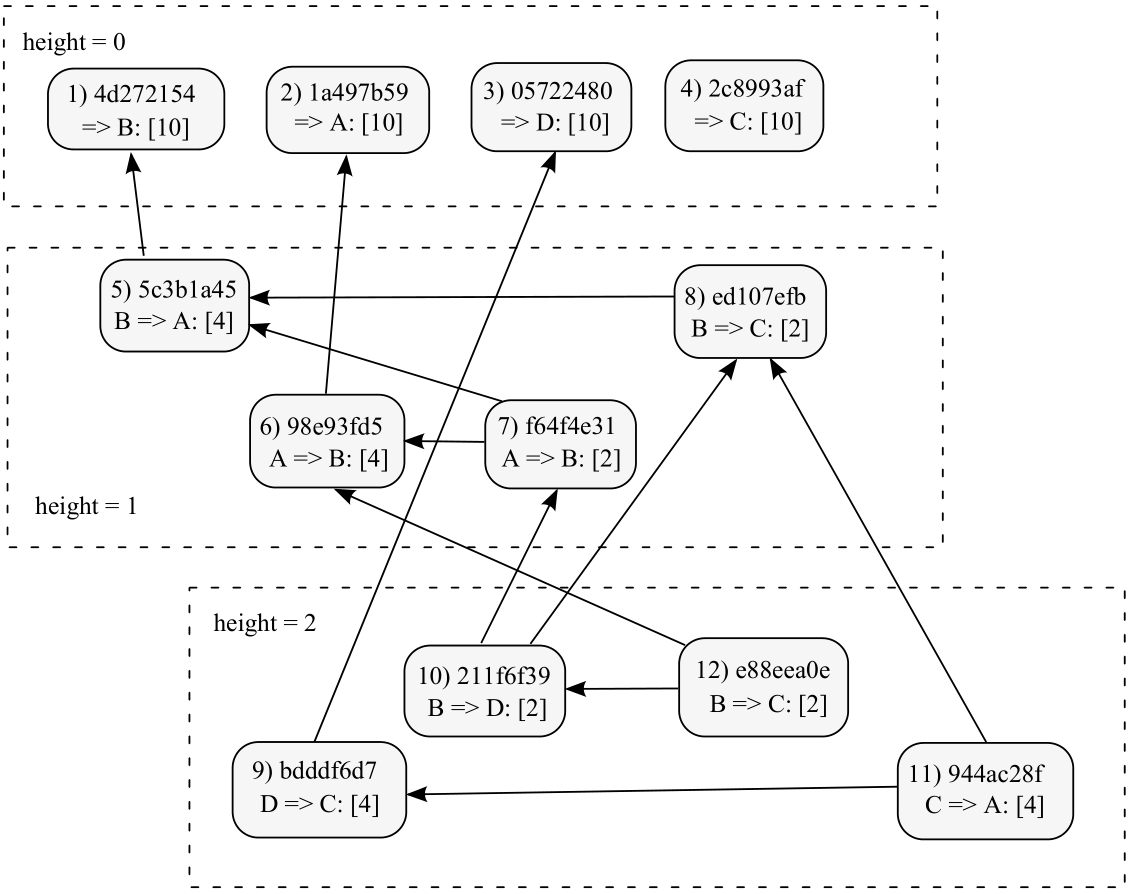}
\end{center}

\subsection{Problem ``Metrical cryptosystem''}

{\bf Solution.} First of all, let us reformulate the problem: the statement ``The cryptosystem presented is correct'' is equivalent to the statement ``It holds $d(x,A)+d(x,B)=d$ for any pair of metrically regular sets $A, B\subseteq \mathbb{F}_2^n$ at the Hamming distance $d$ from each other and any vector $x\in \mathbb{F}_2^n$''.

After considering pairs of metrically regular sets in spaces of small dimension, one may conclude that the cryptosystem is correct, and try to prove it. However, it is not correct. Here we will present the solution of George Beloshapko, Stepan Gatilov, and Anna Taranenko team (Novosibirsk State University, Ledas, Sobolev Institute of Mathematics) as the most elegant.

Consider the Hadamard code of length 16 as $A$:
\begin{gather*}
A=\\
\{(00000000 00000000),\,(00000000 11111111),\\
(00001111 00001111),\,(00001111 11110000),\\
(00110011 00110011),\,(00110011 11001100),\\
(00111100 00111100),\,(00111100 11000011),\\
(01010101 01010101),\,(01010101 10101010),\\
(01011010 01011010),\,(01011010 10100101),\\
(01100110 01100110),\,(01100110 10011001),\\
(01101001 01101001),\,(01101001 10010110) \}.
\end{gather*}

In other words, $A$ is the set of values vectors of all linear functions in 4 variables.
It is easy to check (for example, using computer program), that metrical complement $B$ of the set $A$ is the set of values vectors of all affine function (excluding linear) in 4 variables and vice versa, and the distance between $A$ and $B$ is equal to 8.
\begin{gather*}
B=\\
\{(10010110 01101001),\,(10010110 10010110)\\
(10011001 01100110),\,(10011001 10011001)\\
(10100101 01011010),\,(10100101 10100101)\\
(10101010 01010101),\,(10101010 10101010)\\
(11000011 00111100),\,(11000011 11000011)\\
(11001100 00110011),\,(11001100 11001100)\\
(11110000 00001111),\,(11110000 11110000)\\
(11111111 00000000),\,(11111111 11111111)\}.
\end{gather*}

Now consider vector $x=(00000000 00010111)$.

It is at distance 4 from set $A$ and at distance 6 from set $B$, therefore $$d(x,A)+d(x,B)=10 > 8,$$ which shows us that the cryptosystem is incorrect.

Other solutions include straightforward computer search for metrically regular sets and vectors, for which $d(x,A)+d(x,B)\neq d$. The smallest dimension for which example is found is equal to $7$.
Despite the fact that the cryptosystem is not correct, many participants, who tried to prove that the cryptosystem is correct, received 1 or 2 points for creative ideas.

\subsection{Problem ``Algebraic immunity'' (Unsolved, Special Prize)}
\label{AI}

{\bf Solution.} The problem was completely solved! It was the first time in the Olympiad history. We would like to describe part of the solution proposed by Alexey Udovenko (University of Luxembourg). At first we mentioned that the notion of component algebraic immunity was proposed by C.\,Carlet in \cite{09-Carlet}.

<<It is natural to try simple constructions to build a vectorial Boolean function from
a 1-bit Boolean function. Let $f:\mathbb{F}_2^n\rightarrow\mathbb{F}_2$ be some function such that $AI(f)=\lceil\frac{n}{2}\rceil$.
Let $F_{f,m}:\mathbb{F}_2^n\rightarrow\mathbb{F}_2^m$ be defined as $F_{f,m}(x)=f(x)||f(L(x))||f(L^2(x))||\ldots||f(L^{m-1}(x))$, where $L$ is some linear mapping, for example rotation of the input vector left by one position. We generated random Boolean functions $f$ and for those which had $AI(f)=\lceil\frac{n}{2}\rceil$ we found the largest m such that the corresponding $F_{f,m}$ had high algebraic immunity too, i.\,e.
such that $AI_{comp}(F_{f,m})=\lceil\frac{n}{2}\rceil$.
Note that this approach can produce functions only with $m\leq n$ if $L$ is the rotation
mapping. Here are our results (here $L$ is always rotation left by one):
\begin{itemize}
  \item[$\bullet$] $n=3$: Let $f(x_0,...,x_2)=x_0+x_1+x_1x_2.$ Then $AI_{comp}(F_{f,3})=2$.
  \item[$\bullet$] $n=4$: Let $f(x_0,...,x_3)=x_0x_1x_2+x_0x_1+x_3.$ Then $AI_{comp}(F_{f,4})=2$.
  \item[$\bullet$] $n=5$: Let $f(x_0,...,x_4)=x_0x_1x_2x_3+x_0x_1x_2+x_0x_1x_3+x_0+x_1x_3+x_2x_4+x_4.$ Then $AI_{comp}(F_{f,5})=3$.
  Interestingly, $F_{f,5}$ happens to be a permutation.
  Its lookup table is as follows:
  (0, 24, 12, 20, 6, 29, 10, 1, 3, 23, 30, 26, 5, 22, 16, 19, 17, 9, 27, 2, 15, 21, 13, 7, 18, 4, 11, 14, 8, 28, 25, 31).>>
\end{itemize}

Alexey Udovenko also found a function with optimum component algebraic immunity $\lceil\frac{n}{2}\rceil$ by a rotational symmetries construction for the following values $(n,m)$: (6,6), (7,3), (8,8), (9,2), (10,10). Moreover, he found function with the maximum component algebraic immunity by random search for the $(n,m)$: (4,5), (6,8).

Additionally, he has found $F'$ which is also a permutation of $\mathbb{F}_2^5$ such that $AI_{comp}(F')=3$ but which is differentially 2-uniform (is APN) and its nonlinearity is equal to 12. Therefore, it is more suitable for cryptography. However, the algebraic degree of $F'$ is equal to 3. The lookup table of $F'$ is as follows:
$F'$ = (0, 12, 6, 11, 3, 25, 21, 4, 17, 7, 28, 9, 26, 10, 2, 27, 24, 22, 19, 8, 14, 18, 20, 23, 13, 16, 5, 15, 1, 30, 29, 31).

Alexey Udovenko was the only person who completely solved this problem. We also presented several ideas for finding constructions of vectorial boolean function with optimum component algebraic immunity, but unfortunately it was not completed.

\subsection{Problem ``Big Fermat numbers'' (Unsolved, Special Prize)}

{\bf Solution.} There was no complete solution of this problem. The best solution for this problem was proposed by Alisa Pankova (University of Tartu, Estonia). The main idea of the solution was to show how to use a prime Fermat number $2^{2^k} + 1$ to construct a larger composite Fermat number $2^{2^n} + 1$ for a certain $n > k$. In this way, if there is an arbitrarily large prime Fermat number, then there exists even larger composite Fermat number, so there is no point after which all Fermat numbers become prime. Unfortunately, there was a mistake in the solution.

The team of Vadzim Marchuk, Anna Gusakova, and Yuliya Yarashenia (Belarusian State University) found a very nice euristic bound but the statement was not proven with probability 1. Several interesting attempts were also proposed by Alexey	Udovenko (University of Luxembourg), George Beloshapko, Stepan Gatilov, and Anna Taranenko team (Novosibirsk State University, Ledas, Sobolev Institute of Mathematics), Roman Ginyatullin, Anatoli Makeyev, and Victoriya Vlasova team (Moscow Engineering Physics Institute), Nikolay Altukhov, Vladimir Bushuev, and Roman Chistiakov team (Bauman Moscow State Technical University), Evgeniy Manaka, Aleksandr Sosenko, and Pavel Ivannikov team (Bauman Moscow State Technical University).

\section{Winners of the Olympiad}
\label{winners}

\noindent Here we list information about the winners of NSUCRYPTO-2016 (Tables\;\ref{tab1},\,\ref{tab3},\,\ref{tab2},\,\ref{tab4},\,\ref{tab5}).

\renewcommand{\topfraction}{0}
\renewcommand{\textfraction}{0}

\begin{table}[!h]
\centering\footnotesize
\caption{{\bf Winners of the first round in school section A (``School Student'')}}
\label{tab1}
\renewcommand{\arraystretch}{1.2}
\setlength{\tabcolsep}{1.6mm}
\medskip
\begin{tabular}{|c|l|l|l|c|c|}
  \hline
  Place & Name & Country, City & School & Grade & Sum \\
  \hline
  \hline
  1 & Alexander Grebennikov	& Russia, Saint Petersburg & Presidential PML 239 &	10 & 17 \\
  \hline
  2	& Alexander Dorokhin	& Russia, Saint Petersburg	& Presidential PML 239 &	10 &	14 \\
  \hline
  2	& Ivan Baksheev &	Russia, Novosibirsk	& Gymnasium 6	& 8	& 14 \\
  \hline
  3 &	Vladimir Schavelev &	Russia, Saint Petersburg	& Presidential PML 239 &	10	& 11 \\
  \hline
  3	&	Nikita Dobronravov &	Russia, Novosibirsk	& Lyceum 130 &	11	& 11 \\
  \hline
  Diploma	&	Arkadij Pokazan'ev	& Russia, Novosibirsk	& Gymnasium 6 &	9 &	10 \\
  \hline
  Diploma	&	Arina Prostakova &	Russia, Yekaterinburg	& Gymnasium 94 &	11	& 10 \\
  \hline
  Diploma	&	Gregory Gusev	& Russia, Novosibirsk	& SESC NSU	&11 &	10\\
  \hline
  Diploma	&	Pavel Ivanin &	Belarus, Minsk &	Gymnasium 41 &	9	& 10 \\
  \hline
  Diploma	&	Danil Schrayner &	Russia, Novosibirsk	& Gymnasium 6 &	9	& 9 \\
  \hline
\end{tabular}
\end{table}

\begin{table}[!h]
\centering\footnotesize
\caption{{\bf Winners of the first round, section B (in the category
``Professional'')}}
\label{tab3}
\medskip
\renewcommand{\arraystretch}{1.2}
\renewcommand{\tabcolsep}{1.2mm}
\begin{tabular}{|c|l|p{4.1cm}|p{6.1cm}|c|}
  \hline
  Place & Name & Country, City & Organization & Sum \\
  \hline
  \hline
  1 &	Alexey Udovenko &	Luxembourg, Luxembourg	& University of Luxembourg &	26\\
  \hline
2	&	George Beloshapko	& Russia, Novosibirsk	& Novosibirsk State University &	17\\
  \hline
3	& Ekaterina Kulikova	& Germany, Munich	& -- &	13\\
  \hline
Diploma	& Sergey Belov	& Russia, Obninsk &	Lomonosov Moscow State University	& 9\\
  \hline
Diploma	&	Vadzim Marchuk	& Belarus, Minsk	& Research Institute for Applied Problems of Mathematics and Informatics	& 9\\
  \hline
\end{tabular}
\end{table}

\clearpage
\newpage
\begin{table}[!h]
\centering\footnotesize
\caption{{\bf Winners of the first round, section B (in the category ``University Student'')}}
\label{tab2}
\renewcommand{\arraystretch}{1.2}
\renewcommand{\tabcolsep}{0.8mm}
\medskip
\begin{tabular}{|c|p{2.7cm}|p{2.1cm}|p{3.5cm}|p{4.0cm}|p{0.95cm}|c|}
  \hline
  Place & Name  & Country, City & University & Department & \centering Year & Sum \\
  \hline
  \hline
  1 &	Robert Spencer	& South Africa, Cape Town &	University of Cape Town &	Mathematics and Applied Mathematics &	1 BSc (Hons)	& 24 \\
  \hline
  1	& Henning Seidler	& Germany, Berlin &	Technische Universit\"{a}t Berlin	& Institute of Software Engineering and Theoretical Computer Science	& \centering 6 & 23 \\
  \hline
  2	& Maxim Plushkin & Russia, Moscow	& Lomonosov Moscow State University	& Faculty of Computational Mathematics and Cybernetics	& \centering 1 &	20\\
  \hline
  2 & Pavel Hvoryh & Russia, Omsk &	Оmsk State Technical University	& Information Technologies and Computer Systems Faculty & \centering 5 &	17\\
  \hline
    3	& Irina Slonkina	& Russia, Novosibirsk	& Novosibirsk State University of Economics and Management &	Faculty of Information and Technologies &	\centering 4	& 16\\
  \hline
  3	& Igor Fedorov	& Russia, Novosibirsk &	Novosibirsk State University	&Faculty of Mechanics and Mathematics &	\centering 3 &	14\\
  \hline
  3	& Ivan Emelianenkov &	Russia, Novosibirsk &	Novosibirsk State University &	Mechanics and Mathematics &	\centering 2 &	14\\
  \hline
  3	& Viktoryia Vlasova	& Russia, Moscow	& Moscow Engineering Physics Institute	& Cybernetics and Information Security &	\centering 4 &	13 \\
 \hline
 Diploma &	Pavel Ivannikov &	Russia, Moscow &	Bauman Moscow State Technical University	& Computer Science and Control Systems	& \centering 5 &	12\\
  \hline
  Diploma	& Maria Tarabarina	& Russia, Moscow &	Lomonosov Moscow State University &	Faculty of Mechanics and Mathematics & \centering	3	& 12\\
  \hline
  Diploma	&	Mohamma\-djavad Hajialikhani &	Iran, Tehran	& Sharif University of Technology	& Computer Engineering	& \centering 3	& 12\\
  \hline
  Diploma &	Evgeniy Manaka	& Russia, Moscow &	Bauman Moscow State Technical University	& Computer Science and Control Systems & \centering	5 &	12\\
  \hline
  Diploma & Aleksandr Sosenko &	Russia, Moscow	& Bauman Moscow State Technical University	 Computer & Science and Control Systems	& \centering 5 & 	12\\
  \hline
  Diploma & Vladimir Bushuev &	Russia, Moscow	& Bauman Moscow State Technical University	 Computer & Science and Control Systems &	\centering 5	& 12\\
  \hline
  Diploma	& Roman Lebedev	& Russia, Novosibirsk	& Novosibirsk State University &	Faculty of Physics &	\centering 4 &	11\\
  \hline
  Diploma & Roman Chistiakov	& Russia, Moscow &	Bauman Moscow State Technical University	 Computer & Science and Control Systems & \centering	5	& 11\\
  \hline
  Diploma	& Nikita Odinokih &	Russia, Nоvosibirsk	& Novosibirsk State University	& Faculty of Mechanics and Mathematics	& \centering 4	& 10\\
  \hline
  Diploma	& Alexey Solovev	& Russia, Moscow	& Lomonosov Moscow State University &	Faculty of Computational Mathematics and Cybernetics & \centering	1 &	10\\
  \hline
  Diploma	& Alexander Tkachev &	Russia, Novosibirsk &	Novosibirsk State University &	Information Technology	& \centering 4	& 10\\
  \hline
\end{tabular}
\end{table}

\begin{table}[!h]
\centering\footnotesize
\caption{{\bf Winners of the second round (in the category ``University Student'')}}
\label{tab4}
\renewcommand{\arraystretch}{1.2}
\renewcommand{\tabcolsep}{0.8mm}
\medskip
\begin{tabular}{|c|p{3.7cm}|p{2.1cm}|p{2.7cm}|p{4.0cm}|c|c|}
  \hline
  Place & Names  & Country, City & University & Department & Year & Sum \\
  \hline
  \hline
  1	&	Maxim Plushkin, Ivan Lozinskiy, Alexey Solovev &	Russia, Moscow	& Lomonosov Moscow State University	& Faculty of Computational Mathematics and Cybernetics	& 1 &	53\\
  \hline
  2	& Alexey Ripinen, Oleg Smirnov, Peter Razumovsky	& Russia, Saratov &	Saratov State University	 & Faculty of Computer Science and Information Technologies &	6 &	44\\
  \hline
  3	& Henning Seidler, Katja Stumpp	& Germany, Berlin	& Technische Universit\"{a}t Berlin	& Institute of Software Engineering and Theoretical Computer Science, Mathematics	& 6	& 35\\
  \hline
  3 & 	Roman Ginyatullin, Anatoli Makeyev, Victoriya Vlasova	& Russia, Moscow	& Moscow Engineering Physics Institute	& Faculty of Cybernetics and Information Security &	5,4,4	& 35\\
  \hline
  3	&	Irina Slonkina	& Russia, Novosibirsk &	Novosibirsk State University of Economics and Management	& Information and Technologies	& 4	& 34\\
  \hline
  3 & 	Nikolay Altukhov, Vladimir Bushuev, Roman Chistiakov &	Russia, Moscow, Korolev &	Bauman Moscow State Technical University	& Computer Science and Control Systems &	5	& 33\\
  \hline
    3 &	Oleg Petrakov, Irina Belyaeva, Vladimir Martyshin	& Russia, Moscow &	Moscow Engineering Physics Institute	& Faculty of Cybernetics and Information Security &	5,1,1 &	32\\
  \hline
  3 & Evgeniy Manaka, Aleksandr Sosenko, Pavel Ivannikov	& Russia, Moscow	& Bauman Moscow State Technical University	& Computer Science and Control Systems	& 5	& 32\\
  \hline
  3	&	Aliaksei Ivanin, Oleg Volodko, Konstantin Pavlov	& Belarus, Minsk	& Belarusian State University	& School of Business and Management of Technology of BSU, Faculty of Applied Mathematics and Computer Science &	6	& 31\\
  \hline
  3	& Roman Lebedev, Ilia Koriakin, Vlad Kuzin &	Russia, Novosibirsk &	Novosibirsk State University	& Faculty of Physics	& 4	& 27\\
  \hline
  3	& Roman Taskin, Prokhor Sadkov, Konstantin Kirienko	& Russia, Yekaterinburg &	Ural State University of Railway Transport	& Information security &	4,4,3	& 21\\
  \hline
  Diploma	& Hamid Asadollahi, Mohammadjavad Hajialikhani &	Iran, Tehran	& Shahid Rajaee University, Sharif university of technology	& Electrical Engineering, Computer Engineering	& 3	& 15\\
  \hline
  Diploma	& 	Alexey Miloserdov, Nikita Odinokih &	Russia, Novosibirsk	& Novosibirsk State University	& Mechanics and Mathematics	& 4	& 11\\
  \hline
  Diploma	&	Mikhail Sorokin &	Russia, Moscow	& Moscow Engineering Physics Institute	& Cybernetics and Information Security &	3	& 10\\
  \hline
  Diploma	& Mikhail Kotov, Alexandra Zhukovskaja, Sergey Batunin	& Russia, Tomsk &	Tomsk State University &	Department of Applied Mathematics and Cybernetics &	4,6,3	& 10\\
  \hline
  Diploma	& 	Victoria Ovsyanikova	& Russia, Saint Petersburg	& ITMO University	& Computer Technologies &	6 &	10\\
  \hline
\end{tabular}
\end{table}

\begin{table}[!h]
\centering\footnotesize
\caption{{\bf Winners of the second round (in the category ``Professional'')}}
\label{tab5}
\renewcommand{\arraystretch}{1.2}
\renewcommand{\tabcolsep}{1.0mm}
\medskip
\begin{tabular}{|c|p{4.9cm}|p{2.7cm}|p{5.7cm}|c|}
  \hline
  Place & Names & Country, City & Organization  & Sum \\
  \hline
  \hline
  1	&	Alexey Udovenko &	Luxembourg, Luxembourg	& University of Luxembourg &	83\\
  \hline
  2	& George Beloshapko, Stepan Gatilov, Anna Taranenko	& Russia,  Novosibirsk	& Novosibirsk State University, Ledas, Institute of Mathematics	& 73\\
  \hline
  3	&	Alisa Pankova	& Estonia, Tartu &	University of Tartu &	29 \\
  \hline
  3	&	Dragos Alin Rotaru, Marco Martinoli, Tim Wood &	United Kingdom, Bristol &	University of Bristol &	28 \\
  \hline
  3	& Nguyen Duc, Bui Minh Tien Dat, Quan Doan &	Vietnam, Ho Chi Minh City	& University of Information Technology & 27 \\
  \hline
  3	&	Vadzim Marchuk, Anna Gusakova, Yuliya Yarashenia	& Belarus, Minsk	& Research Institute for Applied Problems of Mathematics and Informatics, Institut of Mathematics, Belarusian State University &	26\\
  \hline
  3	&	Evgeniya Ishchukova, Ekaterina Maro, Dmitry Alekseev	& Russia, Taganrog &	Southern Federal University	& 25\\
  \hline
  3	&	Sergey Titov &	Russia, Yekaterinburg	& Ural State University of Railway Transport	& 21 \\
  \hline
  Diploma &	Philip Lebedev, Mikhail Finoshin, Roman Burmistrov	& Russia, Moscow	& Moscow Engineering Physics Institute	& 16\\
  \hline
  Diploma	&	Sergey Belov, Grigory Sedov	& Russia, Obninsk, Moscow &	Lomonosov Moscow State University &	16\\
  \hline
  Diploma	&	Alisa Koreneva	& Russia, Moscow	& Moscow Engineering Physics Institute	& 12\\
  \hline
\end{tabular}
\end{table}

\FloatBarrier

\section{Photos of the winners}

Here we present photos of all the first place winners in all rounds and categories as well as a photo from the awarding ceremony that was held in December, 2016 in Novosibirsk State University.

\begin{center}
\begin{tabular}{cc}
\includegraphics[width=0.35\textwidth]{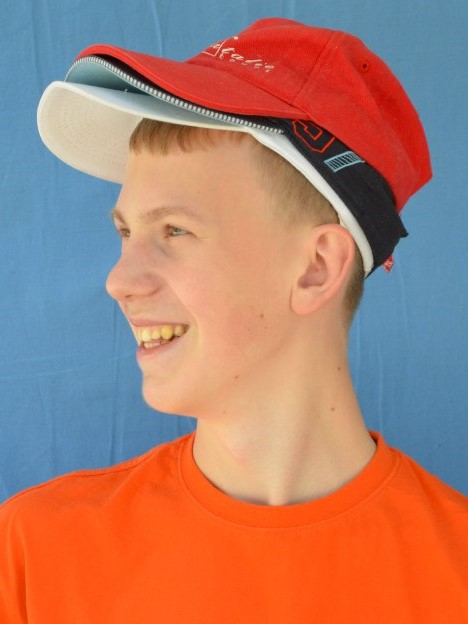}
&

\includegraphics[width=0.35\textwidth]{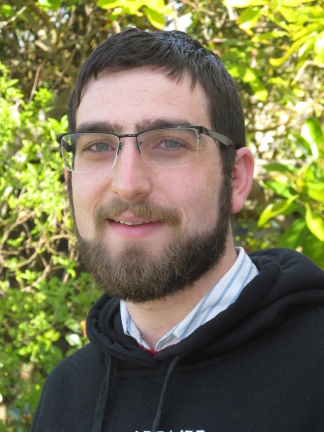}
\\
{\bf Alexander Grebennikov (Russia)}
&
{\bf Robert Spencer (South Africa)}
\\
in category ``School Student'' took
&
in category ``University Student'' took
\\
$\bullet$ the first place in Section A.
&
$\bullet$ the first place in Section B.

\\
 &
\\

\includegraphics[width=0.35\textwidth]{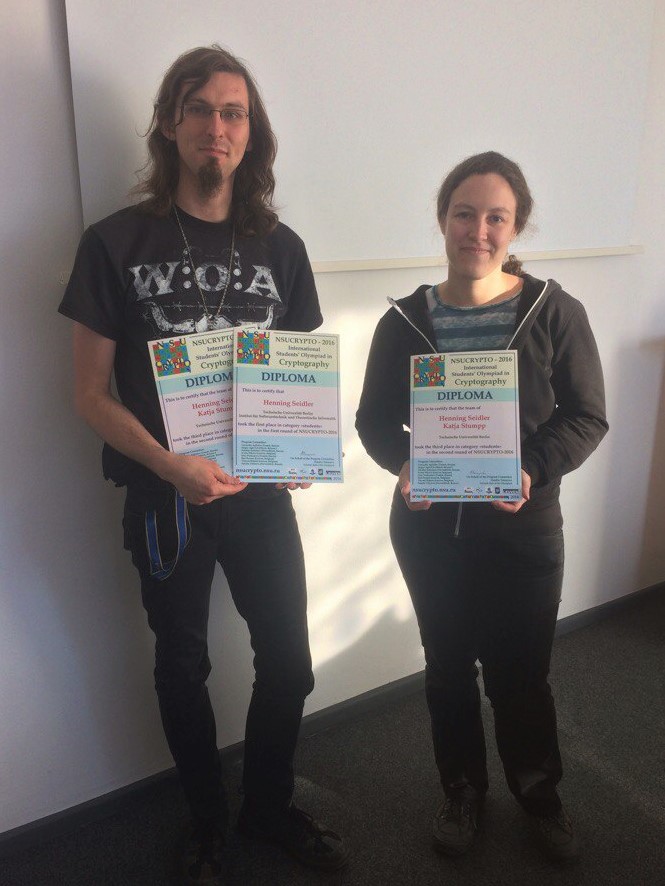}
&

\includegraphics[width=0.35\textwidth]{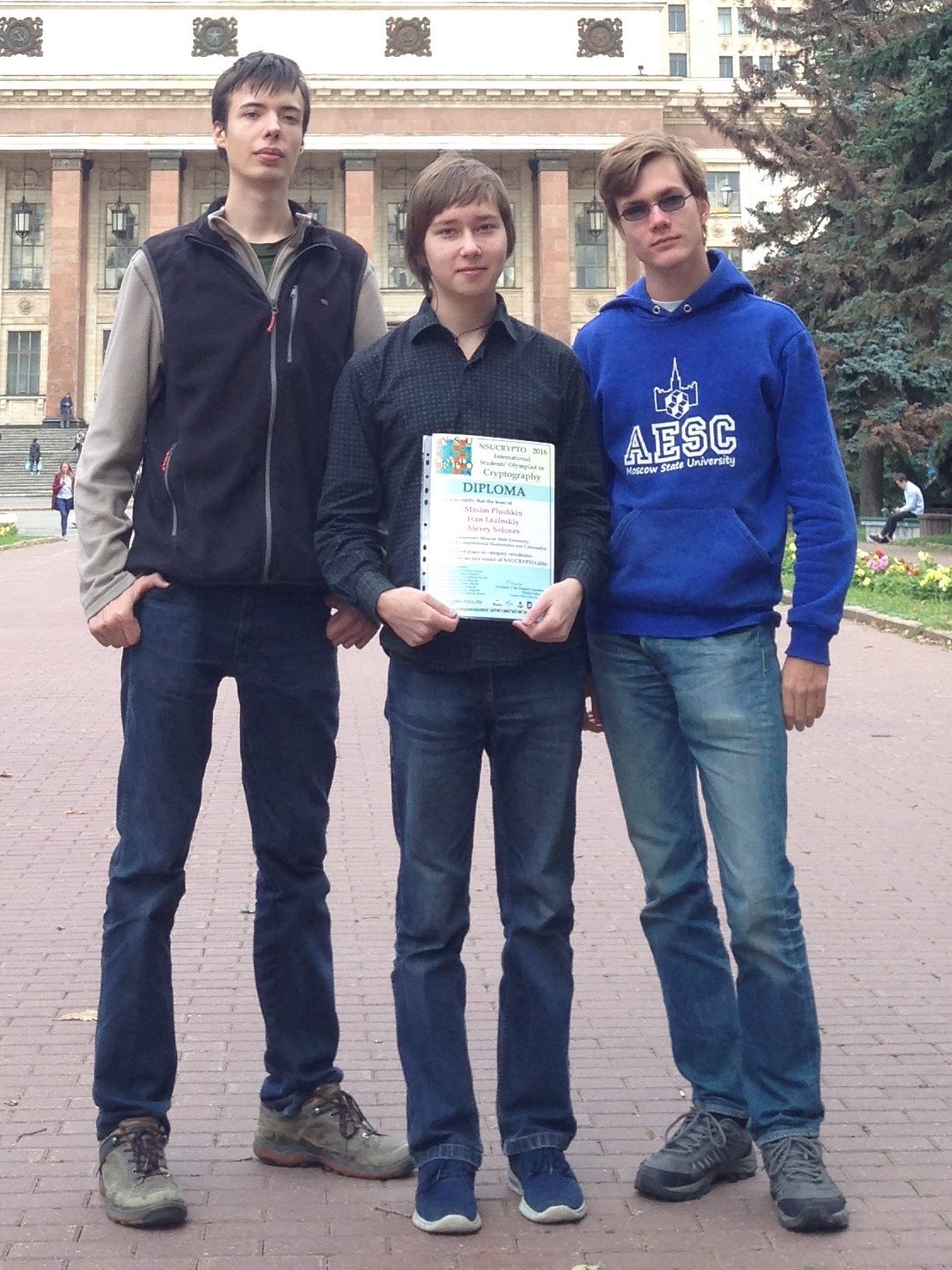}
\\
{\bf Henning Seidler, }
&
{\bf Maxim Plushkin, Ivan Lozinskiy, }
\\
{\bf Katja Stumpp (Germany)}
&
{\bf Alexey Solovev  (Russia)}
\\
in category ``University Student'' took
&
in category ``University Student'' took
\\
$\bullet$ the first place in Section B (H.\;Seidler);
&
$\bullet$ the first place in the second round (team).
\\
$\bullet$ the third place in the second round (team).
&
\\

\end{tabular}
\end{center}

\begin{center}
\begin{tabular}{p{0.55\textwidth}p{0.4\textwidth}}
\

\includegraphics[width=0.55\textwidth]{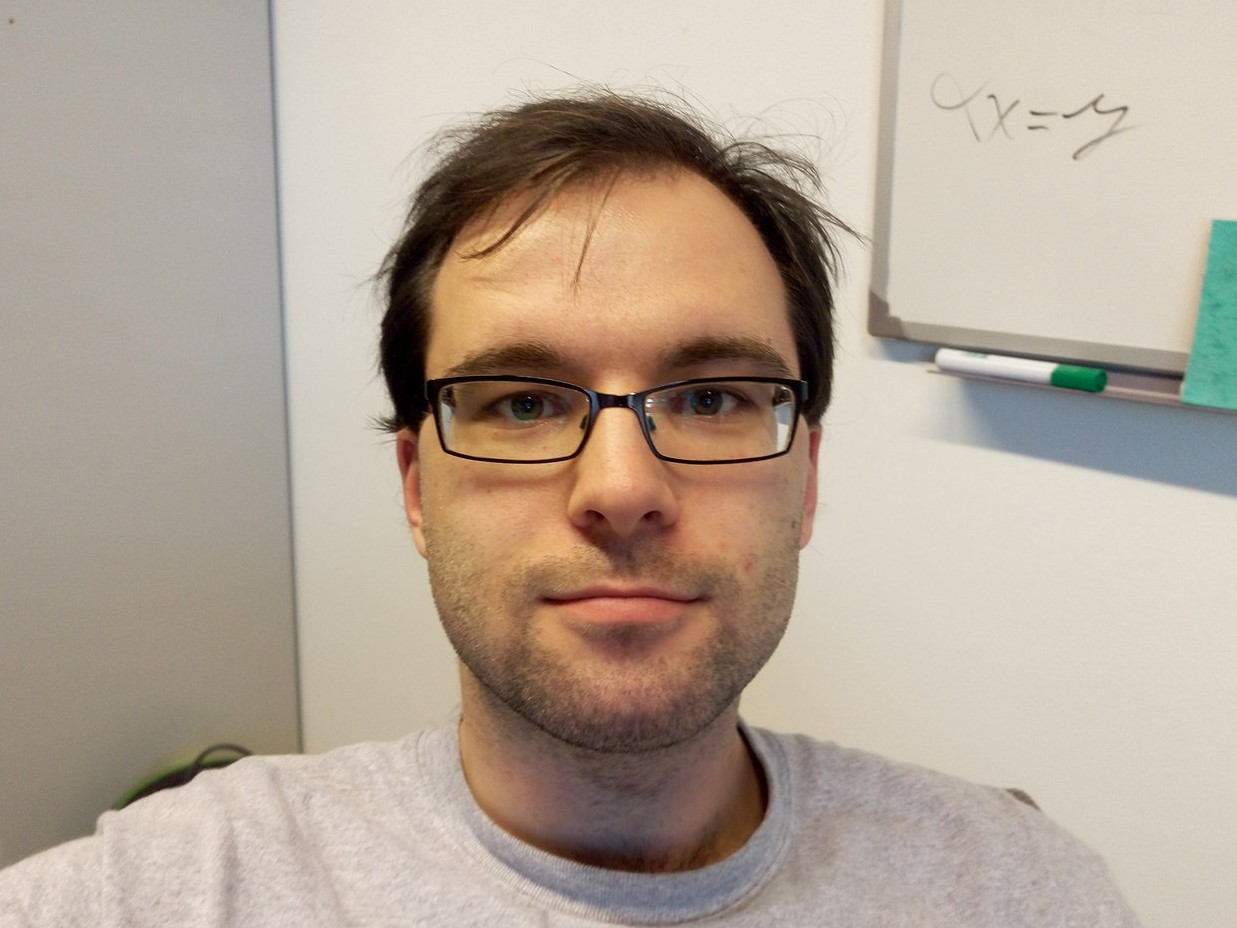}
&
\

\

\

\

\centering
{\bf Alexey Udovenko (Luxembourg)}

in category ``Professional'' took

$\bullet$ the first place in Section B,

$\bullet$ the first place in the second round,
\

and won

$\bullet$ a special prize for solving

the problem  ``Algebraic immunity''.
\end{tabular}
\bigskip
\bigskip

\includegraphics[width=0.97\textwidth]{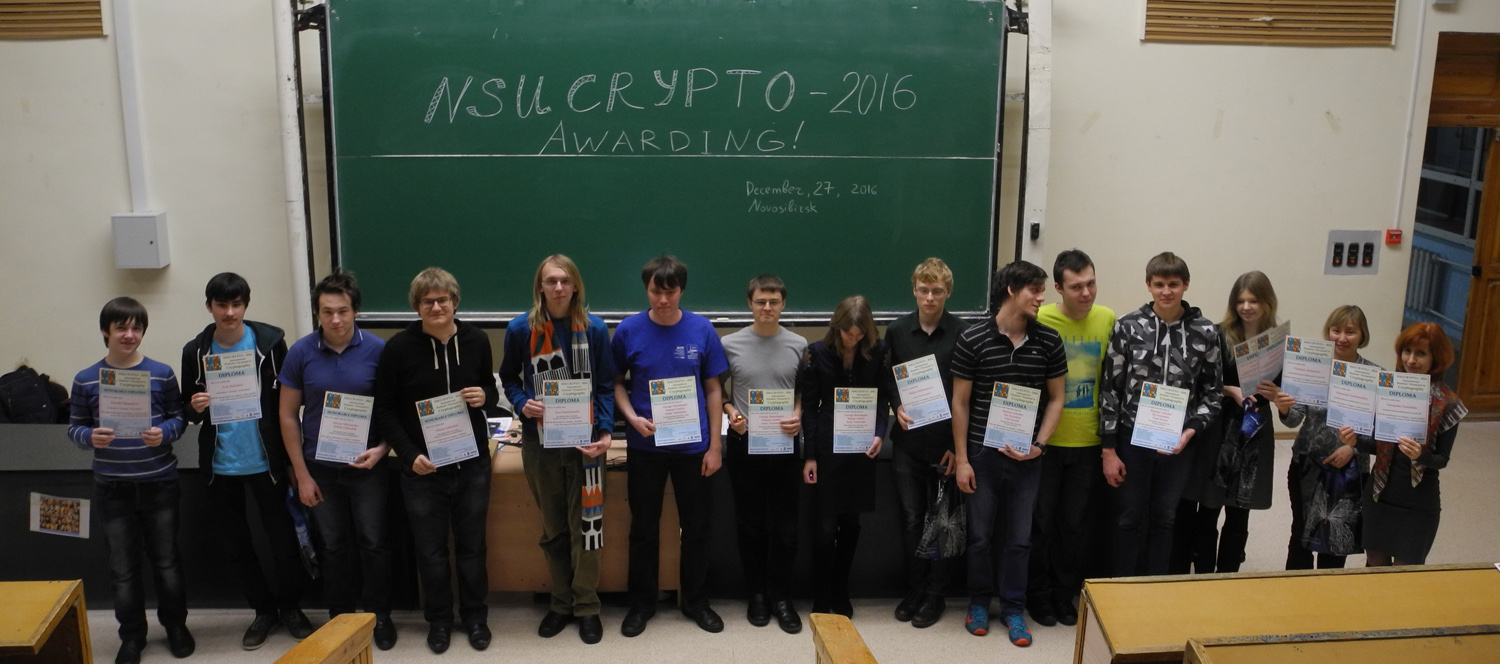}

{\bf Awarding of the winners (Novosibirsk)}
\end{center}

\bigskip

\noindent {\bf Acknowledgements.}
We would like to thank Sergei Kiazhin for his interesting ideas of
the problems. We thank Novosibirsk State University for the
financial support of the Olympiad and invite you to take part in the
next NSUCRYPTO that starts on October 22, 2017. Your ideas on the
mentioned unsolved problems are also very welcome and can be sent to
{\textcolor{blue}{\tt nsucrypto@nsu.ru}}.


\begin{thebibliography}{}

       \bibitem{paper-pdm}
        Agievich\;S., Gorodilova\;A., Kolomeec\;N., Nikova\;S., Preneel\;B.,
        Rijmen\;V., Shushuev\;G., Tokareva\;N., Vitkup\;V.
        Problems, solutions and experience of the first international student's
        Olympiad in cryptography~//
        Prikladnaya Diskretnaya Matematika (Applied Discrete Mathematics). 2015. №\,3, P.\,41--62.

    \bibitem{paper-cryptologia}
        Agievich\;S., Gorodilova\;A., Idrisova\;V., Kolomeec\;N., Shushuev\;G., Tokareva\;N.
        Mathematical problems of the second international student's Olympiad in cryptography~//
        Cryptologia. 2017. Published online.

    \bibitem{09-Carlet}
        Carlet\;C.
        On the Algebraic Immunities and Higher Order Nonlinearities of Vectorial Boolean Functions~//
        in Enhancing Cryptographic Primitives with Techniques from Error Correcting Codes (Proceedings of
        NATO Advanced Research Workshop ACPTECC, Veliko Tarnovo, Bulgaria, October 6–9, 2008) (IOS
        Press, Amsterdam, 2009), pp. 104–116.

    \bibitem{02-DaeRij}
        Daemen\;J., Rijmen\;V.
        The design of Rijndael: AES - the Advanced Encryption Standard
        Springer-Verlag, 2002.

   \bibitem{92-Diffie}
        Diffie\;W., Van Oorschot\;P.\,C., Wiener\;M.\,J.
        Authentication and authenticated key exchanges~//
        Designs, Codes and Cryptography. 1992.  V.\,2. I.\,2. P.\,107--125.

  \bibitem{17-Titov}
        Geut\;K., Kirienko\;K., Sadkov\;P., Taskin\;R., Titov\;S.
        On explicit constructions for solving the problem ``A secret sharing''~//
        Prikladnaya Diskretnaya Matematika. Prilozhenie. 2017. №\,10. P.\,68--70. (in Russian)

\bibitem{09-Nakamoto}
         Nakamoto\;S.
         Bitcoin: a peer-to-peer electronic cash system. 2009. Available at \href{https://bitcoin.org/bitcoin.pdf}{https://bitcoin.org/bitcoin.pdf}

\bibitem{11-Rathgeb}
         Rathgeb\;C., Uhl\;C.
         A survey on biometric cryptosystems and cancelable biometrics~//
         EURASIP Journal on Information Security. 2011. V.\,2011. I.\,1. 
\end{thebibliography}
\end{document}